\documentclass[AMA,LATO1COL]{WileyNJD-v2}

\usepackage{amssymb}		
\usepackage{graphicx}		
\usepackage{hyperref}		
\usepackage{amsmath,amssymb}
\usepackage{booktabs}
\usepackage{tabularx}
\usepackage{multirow}
\usepackage{rotating}
\usepackage{xfrac}
\usepackage{rotating}
\usepackage{float}
\usepackage{subfig}
\usepackage{color}
\usepackage{ulem}
\usepackage{soul}
\usepackage{textcomp}
\usepackage[percent]{overpic}

\graphicspath{{LCPCFigs/}}

\articletype{Article Type}%

\received{26 April 2016}
\revised{6 June 2016}
\accepted{6 June 2016}

\begin{document}

\title{Constrained Power Reference Control for Wind Turbines}

\author[1]{Daniel S. Zalkind*}

\author[1]{Marco M. Nicotra}

\author[1]{Lucy Y. Pao}

\authormark{Zalkind \textsc{et al.}}

\address[1]{\orgdiv{Department of Electrical, Computer, and Energy Engineering}, \orgname{University of Colorado Boulder}, \orgaddress{\state{Boulder, CO 80309}, \country{USA}}}

%

\corres{*Daniel S. Zalkind. \email{dan.zalkind@gmail.com}}


\abstract[Summary]{
The cost of wind energy can be reduced by controlling the power reference of a turbine to increase energy capture, while maintaining load and generator speed constraints.
We apply standard torque and pitch controllers to the direct inputs of the turbine and use their set points to change the power output and reduce generator speed and blade load transients.
A power reference controller increases the power output when conditions are safe and decreases it when problematic transient events are expected.
Transient generator speeds and blade loads are estimated using a gust measure derived from a wind speed estimate.
A hybrid controller decreases the power rating from a maximum allowable power.
Compared to a baseline controller, with a constant power reference, the proposed controller results in generator speeds and blade loads that do not exceed the original limits, increases tower fore-aft damage equivalent loads by 1\%, and increases the annual energy production by 5\%.
}

\keywords{power control, power boost, extreme event control, transient estimation, condition monitoring, design constraints}

\maketitle

\section{Introduction}
Typical wind turbine controllers regulate the generator speed, either for optimal power capture in below-rated operation or to ensure safe generator speeds in above-rated operation.
Designs with these goals result in controllers with adequate and robust enough performance to be ubiquitous as a baseline, but they do not reflect the overall design goals of a wind turbine.

For offshore turbines, with larger capital and balance-of-station costs than onshore turbines, the primary factor to reduce the cost of wind energy is increasing energy capture.~\cite{Stehly2018}
However, most recent research in wind turbine controls is aimed at reducing structural loading on the turbine, which could (a) result in blades that are designed to be lighter if the control design is considered during the blade design, or (b) enable longer lasting structures.
Based on recent studies, a 25\% reduction in blade mass results in a 2.5\% reduction in the levelized cost of energy (LCOE)~\cite{Zalkind2019c}, whereas increasing the annual energy production (AEP) by 10\% can result in nearly a 10\% decrease in the LCOE.~\cite{PAO2021}

When designing the hardware components of a turbine, structural loads are calculated using standard simulations and
either fatigue or the worst-case extreme loads from these simulations are used to design the various turbine components; they are referred to as design-driving loads.
Thus, structural loading on a turbine behaves like a constraint on the design, especially for large, modern rotors, where extreme, or peak, loading drives design.
In two recent system-level design analyses for turbines with blades around 100~m in length, peak loads during extreme turbulence were found to drive blade design.~\cite{Zalkind2019c,Bortolotti2016}

Because power capture is directly related to the cost of energy and structural loading acts like a constraint on design, we design a controller to increase power capture while maintaining the same load and generator speed limits.
Generator speed also behaves like a constraint; when a certain threshold is exceeded, supervisory control initiates a shutdown procedure, which reduces the turbine's availability and ultimately its net AEP.
However, the standard approach to controlling generator speed is still based on regulating the speed to a constant set point, while ensuring that there is an adequate margin between that set point and the shutdown threshold.

\begin{figure}
\hspace{-7pt}
\centering
\includegraphics[scale=.85]{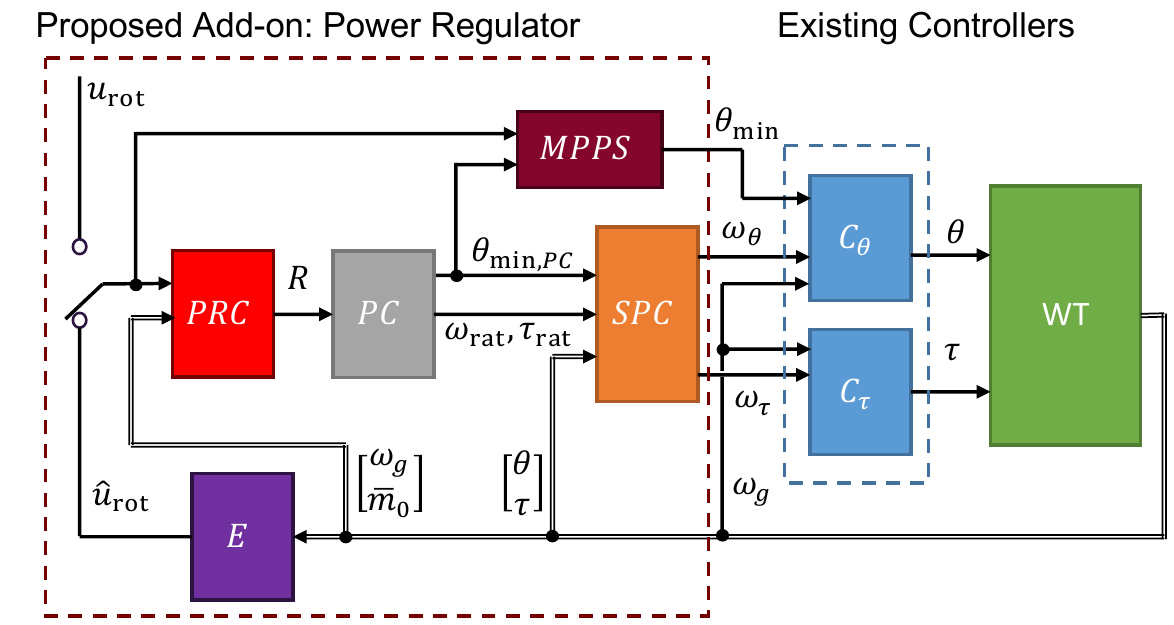}
\caption{High-level schematic of the elements used to control generator power with structural load and generator speed constraints.  The rotor average wind speed is estimated ($\hat{u}_\text{rot}$) and used along with the generator speed $\omega_g$ and filtered collective blade load $\bar{m}_0$ to determine the power reference $R$ in the power reference controller ($PRC$).  The power controller ($PC$) determines the minimum pitch for power control $\theta_{\text{min},PC}$ and rated generator speed $\omega_\text{rat}$ and torque $\tau_\text{rat}$, while the minimum pitch peak shaver (MPPS) determines the lower pitch limit $\theta_\text{min}$ used by the pitch controller $C_\theta$. The set point controller ($SPC$) determines the generator speed set points to the pitch and torque controller ($C_\tau$), $\omega_\theta$ and $\omega_\tau$, respectively. The pitch and torque controllers provide the pitch $\theta$ and torque $\tau$ inputs to the wind turbine (WT).}\label{chCon:fig:overall_bd}
\end{figure}

We implement our controller using these standard, regulator-based control approaches because they are indicative of methods widely used in the field; then we add control modules that change their set points and limits to achieve our goal of increasing power capture and reducing blade loads. 
This modular design approach (shown in Fig.~\ref{chCon:fig:overall_bd}) is consistent with realistic, collaborative controller development, as opposed to replacing existing controllers with a single, monolithic control algorithm to accomplish all tasks. 
A set point controller ensures that both the pitch and torque controller are not simultaneously active, while a power controller uses the minimum pitch angle and rated generator speed to change the power output.
Minimum pitch peak shaving ($MPPS$ in Fig.~\ref{chCon:fig:overall_bd}), or ``thrust clipping,'' is widely referred to in wind turbine control reports~\cite{Hansen2013} as a method of reducing the peak loads that occur near rated wind speeds.
These control elements are commonly used in practice, but are often unpublished or in various, fragmented reports.  
In this article, we present the control laws so that they can all be implemented together by a reader.

In the wind energy industry, controlling generator power is common practice.
For example, the power output is reduced (referred to as de-rating or curtailing) for electrical grid support.~\cite{Aho2012}
Boosting the power output has also been marketed by industrial white papers~\cite{SWP,VWS2014} for increasing the operator's revenue, though manufacturer's methods remain unpublished.


In this article, we seek an answer to the following question: if we actively control the power reference, rather than maintaining it at a static value, how much can we increase power capture while still maintaining safe operation?
An approach, called ``envelope protection,''~\cite{Petrovic2017} uses an optimal control approach to de-rate the turbine only when some limit, or envelope, is expected to be exceeded.
In another approach, power control for reducing peak structural loads was tuned using turbulent statistics and simulation results.~\cite{Kanev2017}
Both approaches inspired this work, which we have extended to include power boosting and peak shaving methods, in what we refer to as power reference control ($PRC$ in Fig.~\ref{chCon:fig:overall_bd}).
We divide the $PRC$ into slow ($PRC^0$) and transient ($PRC^1$) power reference controllers.
In $PRC^0$, the upper limit of the power reference $R$ is determined using a filtered wind speed signal; simulation results guide the design of this component so that power capture is maximized, and generator speed limits are not violated.
$PRC^1$ uses a wind gust measure to estimate transients in the generator speed and blades loads.
During the development of this controller, we found that most over-speed and high load events are caused by a similar wind occurrence: when the wind speed first decreases and then increases.
The proposed gust measure reflects our desire to control this type of event, while a hybrid control scheme de-rates the turbine when estimated transients are expected to exceed some limit.
To the authors' knowledge, the proposed gust measure, transient estimation, and power reference control, with the goal of increasing energy capture, has not been presented in the literature.

Previous versions of this work included more analysis from a control theory perspective, the potential benefits of operating with a known wind input or reduced turbulence, and detailed discussions about the design choices.~\cite{Zalkind2019b,Zalkind2020t}
This article presents the control method in its basic form, using a wind speed estimate for the wind input, in enough detail so that it can be implemented by a wind turbine control designer.
We also include a demonstration of the resulting control signals and a summary of the results that can be achieved when using power reference control for increasing power and constraining peak generator speeds and blade loads.




The various modules are connected as shown in Fig.~\ref{chCon:fig:overall_bd} and the simulation results of various control configurations are presented in Section~\ref{chCon:sec:results}.
The measures used to quantify results are presented in Section~\ref{chCon:sec:metrics} to familiarize the reader with our design goals and procedures.
The open-source wind turbine model is described in Section~\ref{chCon:sec:WT}, which is provided pitch and torque inputs ($\theta$ and $\tau$, respectively) using the proportional-integral (PI) pitch and torque controllers in Section~\ref{chCon:sec:PI}.
The set point controller ($SPC$) and power controller ($PC$) provide generator speed references ($\omega_\theta$ and $\omega_\tau$) given a power reference $R$ as described in Section~\ref{chCon:sec:SPPC}, while the minimum pitch peak shaver ($MPPS$) provides the lower pitch limit $\theta_\text{min}$ to the PI pitch controller using the method in Section~\ref{chCon:sec:MP_PS}.
Finally, the power reference controller ($PRC$) provides the power reference $R$ to the lower levels of control in Section~\ref{chCon:sec:PR}, using turbine states (generator speed $\omega_g$ and filtered collective blade load $\bar{m}_0$) and an estimated wind speed signal ($\hat{u}_\text{rot}$).

\section{Design Measures} \label{chCon:sec:metrics}
We simulate controllers and compare performance in simulation environments specified by a subset of the power producing design load cases (DLCs) determined by the International Electrotechnical Commission:~\cite{Commission2005}
\begin{itemize}
\item DLC 1.2: normal turbulence model (NTM) with 6 seeds at each mean wind speed from cut-in to cut-out (Table~\ref{chCon:tab:turbine}), spaced 2~ms$^{-1}$ apart, and
\item DLC 1.3: extreme turbulence model (ETM), using the same mean wind speeds and number of seeds.
\end{itemize}

\begin{figure}
\centering
\includegraphics[scale=0.85]{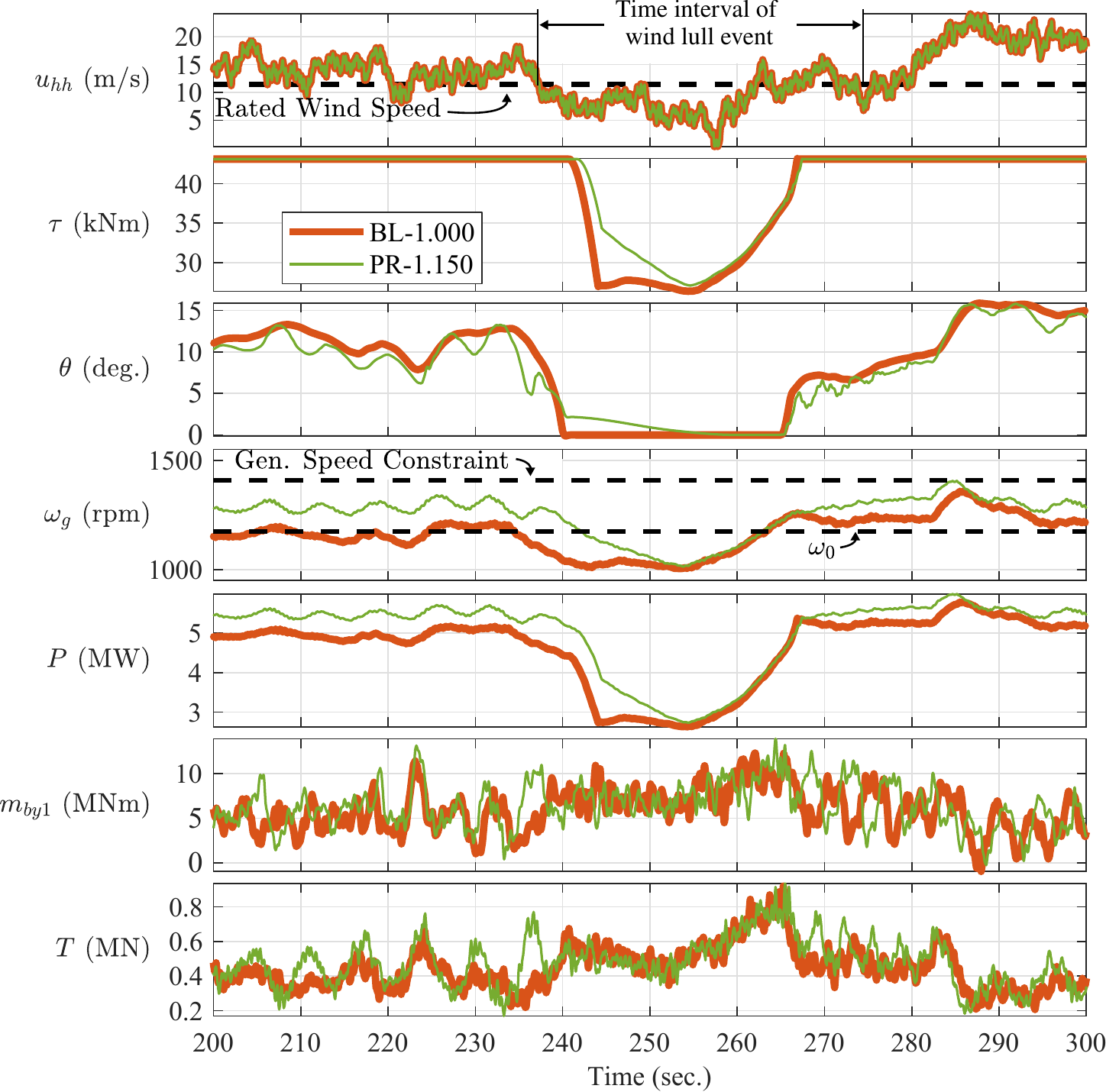}
\caption{Time series of a problematic wind speed lull, where $u_{hh}$, the hub height wind speed, decreases and then increases from about 240--275 sec.  
The generator torque $\tau$  and blade pitch $\theta$ control decrease to regulate the generator speed $\omega_g$ and power $P$.  Low pitch angles with increasing wind speeds lead to large thrust-based loading on the turbine, as indicated by the thrust $T$ and the blade 1 root bending moment $m_{by1}$ around 265 sec.
The baseline control (BL-1.000) is designed to regulate the generator speed to a constant $\omega_0$.
In this article, we propose a controller (PR-1.150) that varies the power reference so that constraints on the generator speed (and loads) are not exceeded, while increasing the power output during safe operation (e.g., before 240 sec.~and after 290 sec.).
} \label{chCon:fig:WT_TS}
\end{figure}

\paragraph{Generator Speed}
Because supervisory wind turbine controllers begin a shutdown procedure when the generator speed exceeds some threshold, we have selected 120\% of the nominal rated generator speed of 1174~rpm (1408~rpm) as the threshold not to exceed; an example time series and the constraint is shown in Fig.~\ref{chCon:fig:WT_TS}.

\paragraph{Extreme Blade Loads}
The blade load that is used to design the structural aspects of a wind turbine blade is referred to as its design load.
Some blade designs use the maximum blade load over all DLC simulations to determine the design load.~\cite{Griffith2014}
Another measure is the characteristic load: it is the maximum (across wind speeds) of the average maximum load (over the turbulent seeds); it has less randomness than the overall maximum.
Blade designers often use the combined (edgewise and flapwise directions) blade load for design, but since the edgewise load is deterministic with the rotor azimuth, the combined load is primarily dependent upon the flapwise load.
For this reason, we focus on the flapwise blade design load in this article.
Peak blade loads typically occur when the pitch angle is low and wind speeds increase, often following a decrease in wind speed; for example, between 240--275 seconds in Fig.~\ref{chCon:fig:WT_TS}.

\paragraph{Rotor Thrust and Tower Fatigue}
Rotor thrust drives tower loading, which is closely related to tower cost.~\cite{Ning2014}
We use the lifetime damage equivalent load (DEL) of the tower base fore-aft (FA) moment to measure the fatigue on the tower; this load tends to increase when the power reference is changed due to extra pitch actuation. 
Load cycles are computed in MLife~\cite{Hayman2012} and extrapolated over the lifetime of the turbine using the wind speed distribution in Table~\ref{chCon:tab:turbine}.
Since rotor thrust and tower loads are closely related to the pitch angle, this measure can also represent the level of pitch actuation.
We try to avoid increasing the tower FA DEL by more than 10\% so that the same tower design can be used.

\paragraph{Energy Capture or Lifetime Average Power} 
The measure with the greatest impact on the cost of wind energy is the annual energy production (AEP).
To compute the AEP, the average power of the six turbulent simulations at each mean wind speed are averaged, resulting in a power curve $P(u)$ for each controller, which is weighted so that
\begin{equation}
AEP = \sum_{u \in U} p(u)P(u), \label{chCon:eq:AEP}
\end{equation}
where $p(u)$ is the (Weibull) wind speed distribution defined in Table~\ref{chCon:tab:turbine}.
Typically, the value in~\eqref{chCon:eq:AEP} is multiplied by the number of hours in the year to determine the annual energy production; we leave this factor off to give a more intuitive measure of the lifetime average power in~MW in the results of Section~\ref{chCon:sec:results}.


\section{Wind Turbine Model} \label{chCon:sec:WT}
We simulate the NREL-5MW reference turbine~\cite{Jonkman2009}, summarized in Table~\ref{chCon:tab:turbine}, using OpenFAST~\cite{OpenFAST} with wind fields generated using TurbSim~\cite{Jonkman2012} according to the design Class 1A.
Because the environmental conditions have a significant effect on the power capture and structural fatigue (via the wind speed distribution), we have chosen to simulate in an environment that represents an offshore site on the east coast of the United States.
\begin{table}[htbp]
\vspace{8pt}
  \centering
  \caption{NREL-5MW reference turbine and environmental parameters}
    \begin{tabular}{lr}
    \textbf{Turbine Parameter} & \textbf{Value} \\
    \midrule
    Rated Power & 5 MW \\
    Rated Rotor Speed & 12.1 rpm \\
    Hub Height & 87 m \\
    Number of Blades & 3 \\
    Rotor Radius & 63 m \\
    Max Chord & 4.65 m \\
    Rotor Position & Upwind \\
    Precone Angle & -2.5 deg. \\
    Baseline Gross Capacity Factor & 45.1\% \\
    Baseline Rated Wind Speed & 11.4~ms$^{-1}$ \\
    Blade Mass & 17.7 Mg \\
          &  \\
    \textbf{Environmental Parameters} &  \\
    \midrule
    Wind Turbine Site Class & Class 1A \\
    Cut-in, cut-out wind speed & 5, 25 ms$^{-1}$ \\
    Mean wind speed at 50 m, hub height & 7.87, 9.11 ms$^{-1}$ \\
    Weibull shape, scale factor & 2.17, 10.3 \\
    \bottomrule
    \end{tabular}%
    \vspace{-8pt}
  \label{chCon:tab:turbine}%
\end{table}%

Increasing the rated generator speed through control actions increases both the capacity factor and the wind speed at which the rated generator power is reached (rated wind speed), while decreasing the rated generator speed has the opposite effect.
In Table~\ref{chCon:tab:turbine}, we show the nominal values without power regulation.
Using the wind speed distribution in Table~\ref{chCon:tab:turbine} without power regulation results in a capacity factor of 45.1\%; this capacity factor is less than more recently developed turbines, which have larger rotors relative to the rated generator power.
However, this turbine represents a realistic case for using the power boosting functionality presented in this article as an aftermarket addition to existing infrastructure.

\section{PI Torque and Pitch Control} \label{chCon:sec:PI}

\begin{figure}
\centering
\includegraphics[scale=0.85]{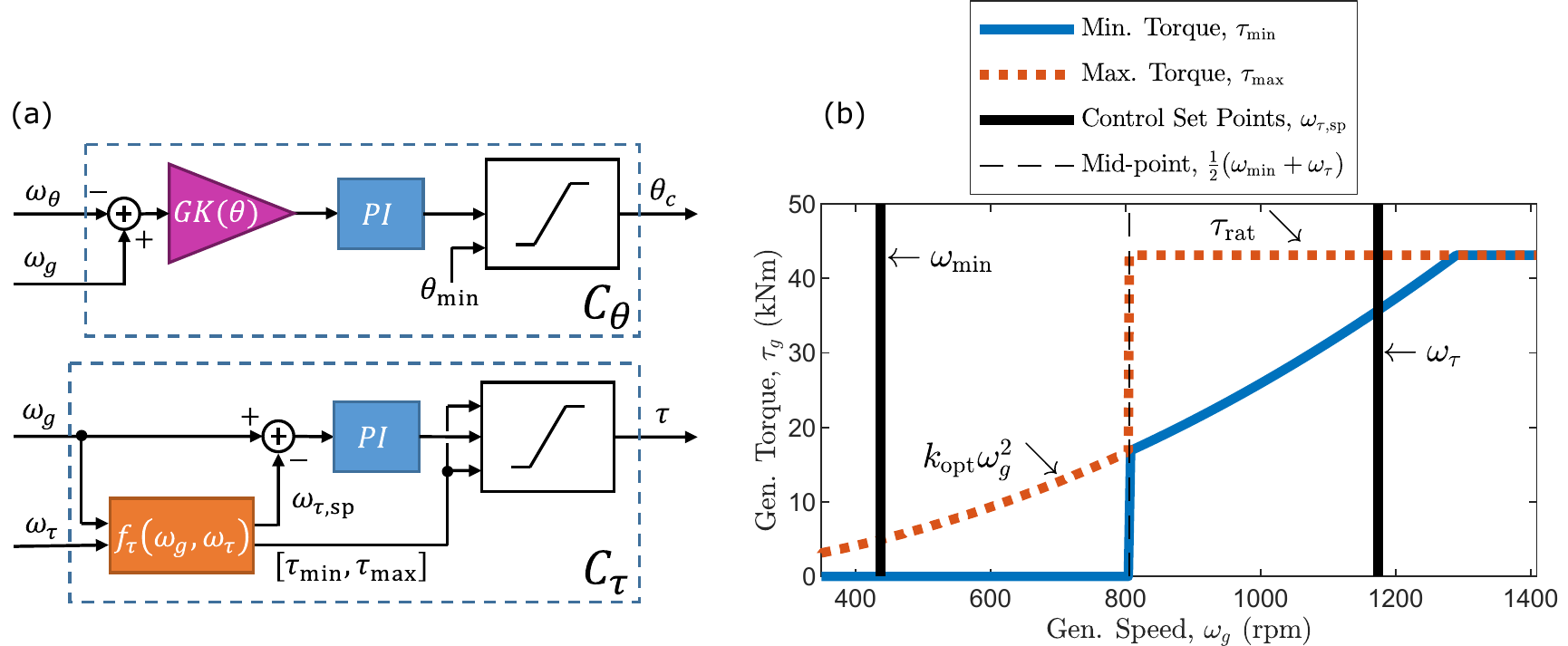}
\caption{PI control modules (a) provide the torque $\tau$ and pitch $\theta$ controls to the wind turbine, subject to upper and lower limits on torque and pitch. Note that $\theta_c$ is the commanded pitch input to the pitch actuator, $\theta$ is the current pitch angle of the blades, and $GK(\theta)$ is a gain scheduling parameter.  The nonlinear function $f_\tau(\omega_g,\omega_\tau)$ given in~\eqref{chCon:eq:torqSp} is visualized in (b), which determines the generator speed set point $\omega_{\tau,\text{sp}}$ and torque saturation limits $[\tau_\text{min},\tau_\text{max}]$, depending on whether the generator speed is near cut-in or rated operation. Anti-windup schemes are necessary for each integrator in the proportional-integral (PI) controllers, which are both active at all times, but may be saturated.
}\label{chCon:fig:PI_control}
\end{figure}

To control the direct inputs to the wind turbine, torque $\tau$ and pitch $\theta$, we use proportional-integral (PI) controllers (Fig.~\ref{chCon:fig:PI_control}), which are commonly used control architectures, both in research and industry, and we would like to maintain their use when designing the other control modules.

\subsection{Pitch Control}
The pitch controller is a gain-scheduled PI control.
Because the magnitude of the sensitivity of power to pitch angle decreases as the pitch increases,~\cite{Jonkman2009} the PI gains of the pitch controller should decrease to maintain consistent generator speed regulation.
The gain-correction factor, $GK(\theta)$ in Fig.~\ref{chCon:fig:PI_control}(a),
is applied to the generator speed error, rather than to the gains themselves for a simpler implementation and to maintain appropriate gains throughout above-rated wind speeds.\cite{Dunne2016}

The PI gains $k_{P,\theta}$ and $k_{I,\theta}$ in Table~\ref{chCon:tab:PIparams} are determined using the desired natural frequency (or bandwidth) $\omega_{\text{reg},\theta}$  and the damping ratio $\zeta_{\text{reg},\theta}$ of the generator speed regulator mode.\cite{Jonkman2009}
Increasing the bandwidth $\omega_{\text{reg},\theta}$ results in a faster pitch response to wind disturbances and better generator speed regulation, but also greater structural loading.
We determine the regulator mode using a data-driven optimization procedure with the goal of reducing fatigue loads on the tower, such that no maximum generator speed transients exceed 120\% of the nominal rated generator speed.\cite{Zalkind2020a}
Since the worst case generator overspeed occurs during an ETM simulation with a 20~ms$^{-1}$ mean wind speed (Fig.~\ref{chCon:fig:ResGen_BL}), that wind input is used to tune the parameters of the pitch controller.
The optimization procedure results in a regulator mode with a lower bandwidth and higher damping ratio than the prescribed values originally recommended for this rotor and pitch controller.\cite{Jonkman2009}
The reduced bandwidth is also beneficial when the pitch control set point is changed using the transient power reference control (Section~\ref{chCon:sec:PR_Stab}).
The pitch command $\theta_c$ is filtered using a second-order butterworth filter with a cutoff frequency of 1~Hz, which represents the pitch actuator, to determine the pitch $\theta$ of the turbine.

The pitch control set point $\omega_\theta$ is normally the nominal rated generator speed ($\omega_0$=1174~rpm), but we control this value to increase power capture or decrease generator speed transients, by increasing or decreasing $\omega_\theta$, respectively.
The minimum pitch setting $\theta_\text{min}$ is usually the optimal pitch angle of the blades with respect to energy capture, but we control this value to change the power output in Section~\ref{chCon:sec:PR0} and reduce peak blade loading in Section~\ref{chCon:sec:MP_PS}.

\subsection{Torque Control}
We implement a PI torque controller similar to previous work,\cite{Bossanyi2003} which provides a smoother torque control and a simpler implementation than lookup-table-based control schemes if we change the rated generator speed. 
The torque control gains $k_{P,\tau}$ and $k_{I,\tau}$ in Table~\ref{chCon:tab:PIparams} are derived using a similar approach to the pitch control gains by defining a ``regulator mode'' of the torque controller, which has been tuned to balance energy capture with power fluctuations.~\cite{Zalkind2020t}

Depending on whether the turbine is near cut-in or rated operation, the generator speed set point $\omega_{\tau,\text{sp}}$ and torque limits $[\tau_\text{min},\tau_\text{max}]$ are changed 
using the nonlinear function
\begin{equation}
[\omega_{\tau,\text{sp}},\tau_\text{min},\tau_\text{max}]^T = 
f_\tau(\omega_g,\omega_\tau) = 
\begin{cases}
[\omega_\text{min},0,k_\text{opt}\omega_g^2]
& \text{if}\quad\omega_g < \frac{1}{2}(\omega_\text{min} + \omega_\tau),  \\
[\omega_{\tau} , k_\text{opt}\omega_g^2 , \tau_\text{rat}]
& \text{otherwise,}
\end{cases}\label{chCon:eq:torqSp}
\end{equation}
illustrated in Fig.~\ref{chCon:fig:PI_control}(b).

\begin{table}[htbp]
  \centering
  \caption{Parameters for proportional-integral pitch and torque controllers.}
    \begin{tabular}{rlcr}
    \toprule
          & Parameter & Variable & Value \\
    \midrule
          & Total Drivetrain Inertia & $J_\text{tot}$  & 4.38$\times\text{10}^7~\text{kg m}^2$ \\
    \multicolumn{1}{l}{Turbine} & Minimum Gen. Speed & $\omega_\text{min}$  & 436.5 rpm \\
    \multicolumn{1}{l}{Parameters} & Nominal Rated Gen. Speed & $\omega_0$  & 1174 rpm \\
          & Gearbox Ratio & $G$     & 97 \\
    \midrule
          & Gain Scheduling Param. & $\theta_k$ & 4.71 deg. \\
    \multicolumn{1}{l}{Pitch} & Pitch Regulator Mode: Natural Frequency & $\omega_{\text{reg},\theta}$ & 0.275 rad s$^{-1}$ \\
    \multicolumn{1}{l}{Control} & Pitch Regulator Mode: Damping Ratio & $\zeta_{\text{reg},\theta}$ & 1.59$^*$ \\
    \multicolumn{1}{l}{Parameters} & Pitch Regulator: Proportional Gain & $k_{P,\theta}$ & 0.0143 sec. \\
          & Pitch Regulator: Integral Gain & $k_{I,\theta}$ & 7.18$\times\text{10}^{-4}$ \\
    \midrule
    \multicolumn{1}{l}{Torque}      & Optimal Torque Control Gain & $k_\text{opt}$  & 0.22 $\text{N m rpm}^{-2}$ \\
    \multicolumn{1}{l}{Control} & Torque Regulator: Proportional Gain & $k_{P,\tau}$ & 9.75 N m s rad$^{-1}$ \\
    \multicolumn{1}{l}{Parameters} & Torque Regulator: Integral Gain & $k_{I,\tau}$ & 4.88 N m rad$^{-1}$\\
    \bottomrule
    \multicolumn{4}{l}{$^*$A damping ratio greater than 1 results in two real poles, rather than two complex conjugate poles.}
    \end{tabular}%
    
  \label{chCon:tab:PIparams}%
\end{table}%




\section{Set Point and Power Control} \label{chCon:sec:SPPC}

\subsection{Set Point Control} \label{chCon:sec:SSC}
To ensure that both torque and pitch are not simultaneously controlling the generator speed, we use a set point controller (SPC) developed by Sowento.\cite{Sowento}
If both PI torque and pitch controllers have the same generator speed set points, both will be active, leading to poor performance in terms of power production and increased pitch actuation, which increases the loading on the turbine.
\begin{table}
  \begin{minipage}{0.425\textwidth}
            \centering
\includegraphics[scale=0.75]{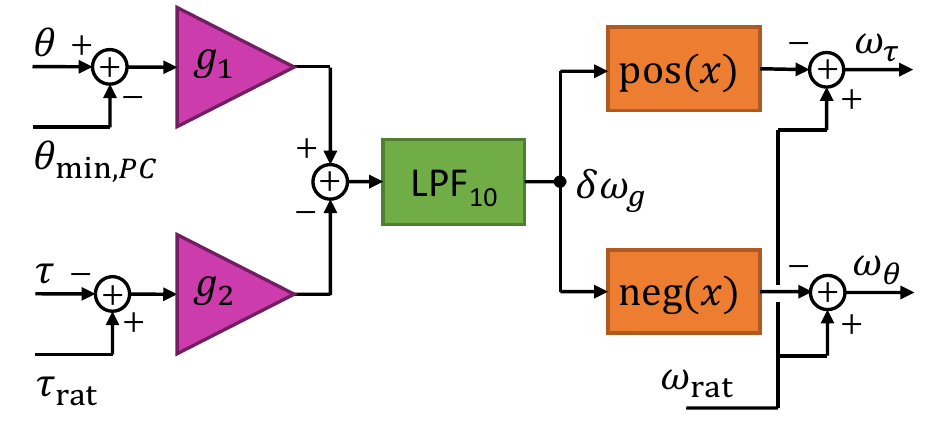}
\captionof{figure}{Set point smoothing control used to determine generator speed set points $\omega_\theta$ and $\omega_\tau$ for PI-control in Fig.~\ref{chCon:fig:PI_control}, given the current pitch angle $\theta$, the minimum pitch setting for power control $\theta_{\text{min},PC}$, torque $\tau$, rated torque $\tau_\text{rat}$, and rated generator speed $\omega_\text{rat}$.}\label{chCon:fig:SSC}
  \label{tab:v2_struct2}
    \end{minipage}%
    \hfill
    \begin{minipage}{0.55\textwidth}
  \centering
  \caption{set point and power control parameters.} \label{chCon:tab:SPC}
    \begin{tabular}{llcr}
    \toprule
          & Parameter & Variable & Value \\
    \midrule
    Turbine & Rated Gen. Torque & $\tau_\text{rat}$  & 43.1 kNm \\
    Parameters & Fine Pitch Angle & $\theta_\text{fine}$ & 0 deg. \\
    \midrule
    Set point Control & Torque Set point Bias & $g_1$    & 33.3 rpm deg.$^{-1}$ \\
    Design Choices & Pitch Set point Bias & $g_2$    &  2.79 rpm (kNm)$^{-1}$ \\
    \bottomrule
    \end{tabular}%
    \end{minipage}
\end{table}

The SPC module, shown in Fig.~\ref{chCon:fig:SSC}, uses a generator speed set point bias,
\begin{equation}
\delta \omega_g = \text{LPF}_{10}  \left\{ g_1 (\theta - \theta_{\text{min},PC}) - g_2 (\tau_\text{rat} - \tau) \right\},
\end{equation}
where the gains $g_1$ and $g_2$ are both positive and $\text{LPF}_{10}$ is a low-pass filter
\begin{equation}
\text{LPF}_{\tau_l} = \frac{(\sfrac{2\pi}{\tau_l})^2}{s^2 + \left(\sfrac{2\pi\sqrt{2}}{\tau_l}\right)s + \left(\sfrac{2\pi}{\tau_l}\right)^2}, \label{chCon:eq:LPF2}
\end{equation}
where $\tau_l = 10$~sec. and $s = j\omega$ is the complex frequency; $\tau_l$ was chosen to ensure a smooth $\delta \omega_g$ signal.
The set point bias is applied to the torque and pitch set points, depending upon the sign of $\delta\omega_g$:
\begin{equation}
\omega_\tau =
\begin{cases}
\omega_\text{rat} - \delta \omega_g & \text{if}\quad \delta\omega_g > 0 \\
\omega_\text{rat} & \text{otherwise}
\end{cases} \label{chCon:eq:tau_sp_pw}
\end{equation}
and 
\begin{equation}
\omega_\theta =
\begin{cases}
\omega_\text{rat} - \delta \omega_g & \text{if}\quad \delta\omega_g < 0 \\
\omega_\text{rat}  & \text{otherwise},
\end{cases} \label{chCon:eq:theta_sp_pw}
\end{equation}
which is related to whether the turbine is in above- or below-rated operation.
Typically, $\delta\omega_g > 0$ during above-rated operation and vice versa.
The nonlinear functions in~\eqref{chCon:eq:tau_sp_pw} and~\eqref{chCon:eq:theta_sp_pw} are represented in the block diagram in Fig.~\ref{chCon:fig:SSC}, where
\begin{equation}
\text{pos}(x) =
\begin{cases}
x & \text{if}\quad x > 0 \\
0  & \text{otherwise}
\end{cases} \label{chCon:eq:tau_sp_bd}
\end{equation}
and
\begin{equation}
\text{neg}(x) =
\begin{cases}
x & \text{if}\quad x < 0 \\
0  & \text{otherwise}.
\end{cases}\label{chCon:eq:theta_sp_bd}
\end{equation}

During above-rated operation, $\theta > \theta_{\text{min},PC}$, which increases $\delta\omega_g$ and reduces $\omega_\tau$ in~\eqref{chCon:eq:tau_sp_pw}, biasing the torque control towards rated torque $\tau_\text{rat}$.
During below-rated operation, $\tau < \tau_\text{rat}$, reducing $\delta\omega_g$, which, if negative, increases $\omega_\theta$ in~\eqref{chCon:eq:theta_sp_pw} and biases the pitch control towards its minimum saturation limit $\theta_{\text{min},PC}$.
The design choices for the gains in this controller are shown in Table~\ref{chCon:tab:SPC} and they are tuned based on the power capture and power variation of near-rated simulations.~\cite{Zalkind2020t}

\subsection{Power Controller} \label{chCon:sec:PC}
To change the power output of the turbine, the rated generator speed is controlled so that
\begin{equation}
\omega_\text{rat} = R \omega_{0},
\end{equation}
where $\omega_{0}$ is the original rated generator speed (1174~rpm) and $R$ is the power reference factor; this works across all operating wind speeds.
\begin{figure}
\centering
\raisebox{17pt}{\includegraphics[width=.375\linewidth]{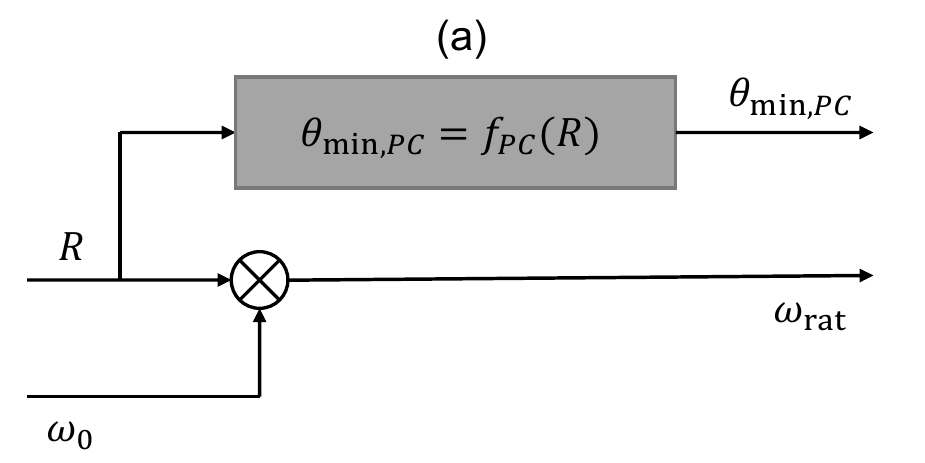}}
\hspace{1 cm}
\includegraphics[width=.325\linewidth]{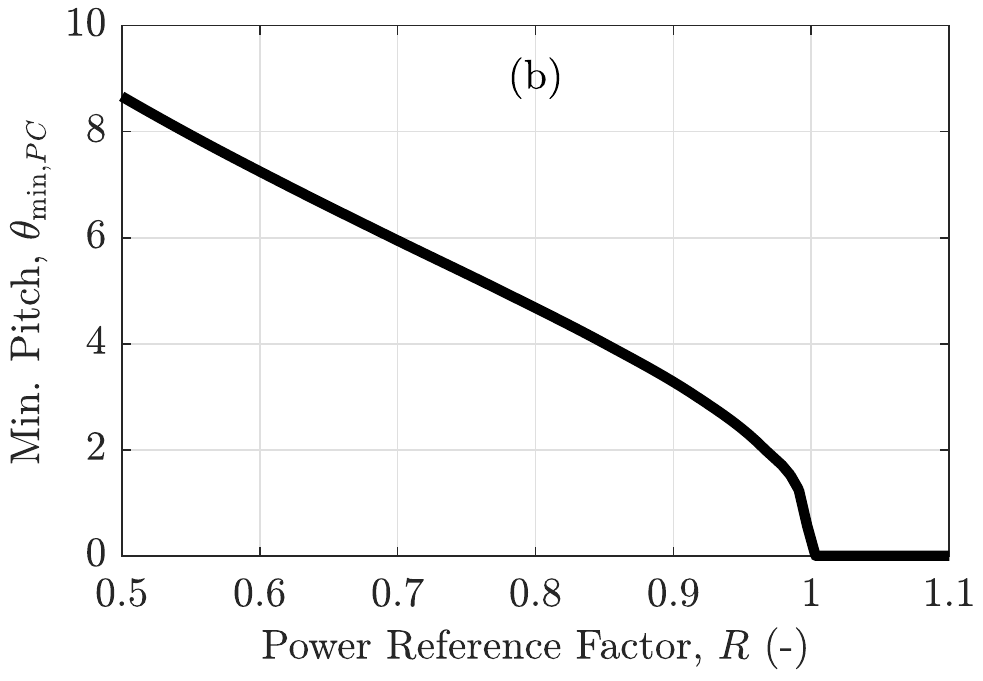}
\caption{The power controller (a), given a power reference factor $R$, sets the minimum pitch angle $\theta_{\text{min},PC}$ according to $f_{PC}(R)$ as shown in (b) and the rated generator speed $\omega_\text{rat}$. 
}\label{chCon:fig:PC}
\end{figure}

To de-rate the turbine in below-rated operation, the minimum pitch setting is controlled using $\theta_{\text{min},PC} = f_{PC}(R)$, which, along with the minimum pitch peak shaver in Section~\ref{chCon:sec:MP_PS}, contributes to the saturation limit used by the PI pitch controller.
With a minimum pitch setting greater than the optimal (or fine) pitch setting $\theta_\text{fine}$, the aerodynamic torque and generator speed are reduced. 
The function $f_{PC}(R)$ that determines the minimum pitch setting is shown in Fig.~\ref{chCon:fig:PC}(b); it is determined
using simulations with a below-rated, constant wind inflow by varying the minimum pitch angle, finding the power output compared to the optimal output, and inverting the function.~\cite{Zalkind2020t}

We cannot always increase power output above the rated value.
However, we can always de-rate the turbine, which is desirable whenever a problematic gust event occurs.
Going forward, we use the control architecture described to this point, with $R = 1$, as a baseline for comparison.


\section{Peak Shaving Using Minimum Pitch Control} \label{chCon:sec:MP_PS}


The goal of the minimum pitch peak shaving (MPPS) controller is to prevent instances where there is a low pitch angle and high wind speed, causing a large torque and thrust on the rotor.
In the simulations performed using the previously described controller (with a constant power reference factor of $R = 1$), this occurs during wind lulls at high wind speeds.
The MPPS control determines the lower limit on the pitch command (Fig.~\ref{chCon:fig:PI_control}(a)):
\begin{equation}
\theta_\text{min} = \max \{ \theta_{\text{min},PC}, \theta_{\text{min},PS} \} \label{chCon:eq:theta_min}
\end{equation}
where $\theta_{\text{min},PC}$ is the minimum pitch setting for power control (Section~\ref{chCon:sec:PC}) and $\theta_{\text{min},PS}$ is the minimum pitch for peak shaving. 


%
%
%

\begin{figure}
\centering
\vspace{3pt}
\includegraphics[width=0.485\linewidth]{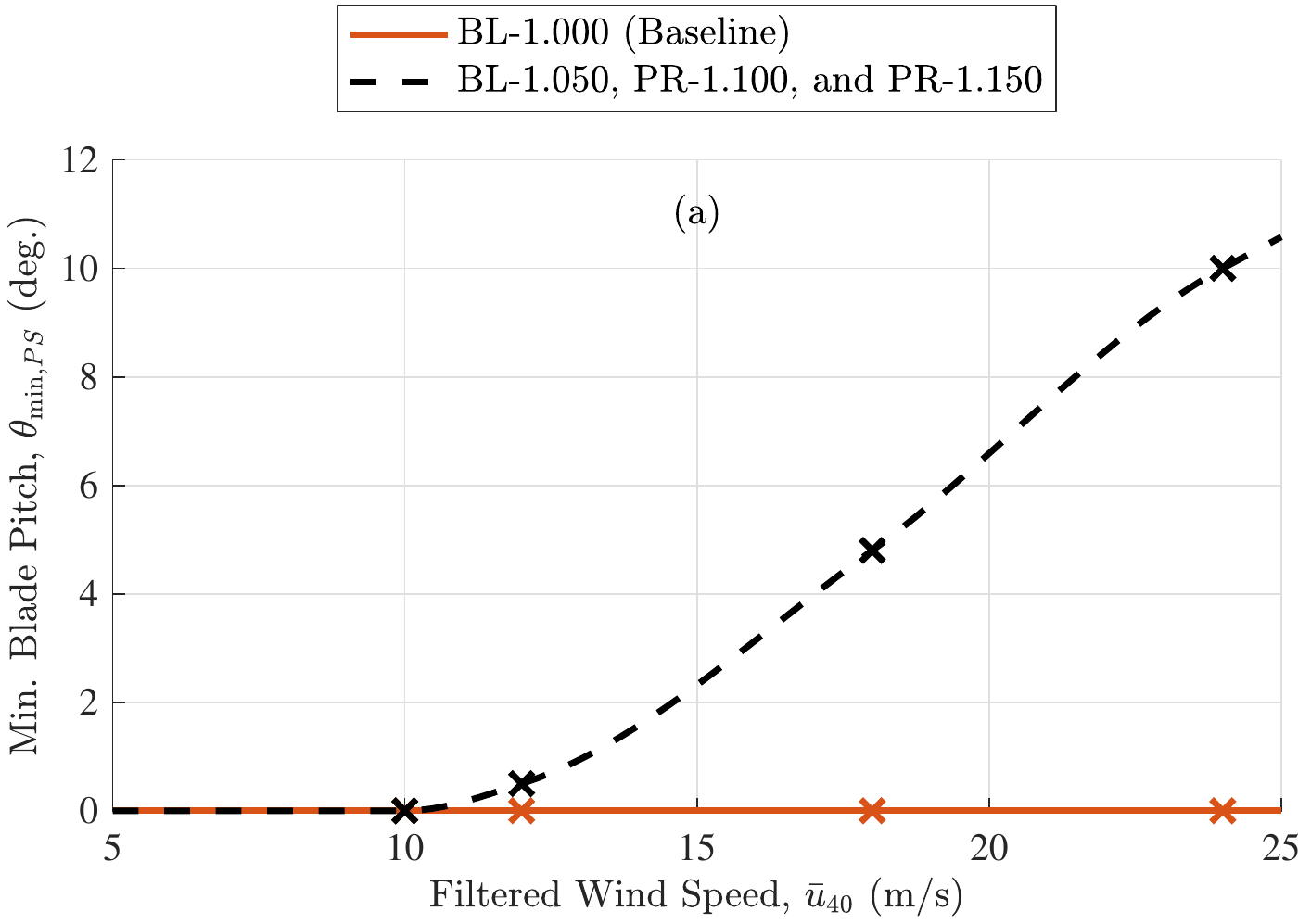}
\hfill
\includegraphics[width=0.485\linewidth]{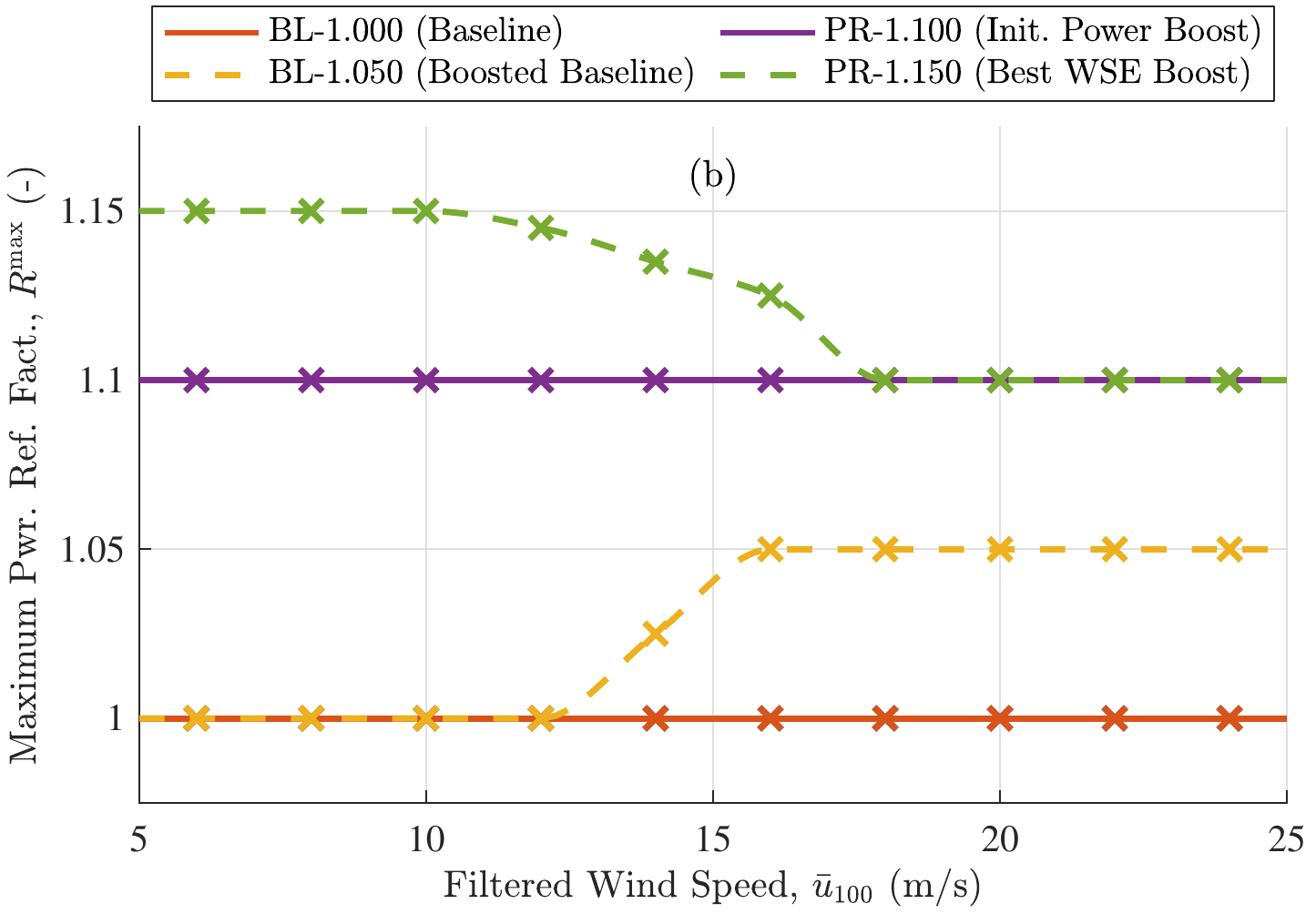}
\caption{Minimum pitch peak shaving control (a) and maximum power reference factor (b) for the various control cases described in Section~\ref{chCon:sec:results}.}\label{chCon:fig:MPPS}
\end{figure}

A lookup table defines the function $\theta_{\text{min},PS}(\bar{u}_{40})$, which depends on the slow low-pass filtered wind speed estimate $\bar{u}_{40} = \text{LPF}_{40} \big\{ \hat{u}_\text{rot}\big\}$, where $\text{LPF}_{\tau_l}$ is defined in~\eqref{chCon:eq:LPF2}.
A $\tau_l$ of 40~sec.~ensures that the minimum pitch signal does not change too rapidly, introducing dynamics in near rated wind speeds, and that there is some memory of the mean wind speed, so that when problematic wind lulls occur, the minimum pitch value is still high enough to avoid large rotor thrusts.
Between the breakpoints in Fig.~\ref{chCon:fig:MPPS}(a), cubic interpolation determines  $\theta_{\text{min},PS}$.  
We use a break point at 10~ms$^{-1}$, just below rated, such that $\theta_{\text{min},PS}(10) = \theta_\text{fine}$.
The minimum blade pitch $\theta_{\text{min},PS}$ at 12~ms$^{-1}$, just above the rated wind speed, is tuned to trade off power capture and peak loading.~\cite{Zalkind2020a}
The minimum pitch angles at the high wind speed breakpoints (18 and 24~ms$^{-1}$) are chosen so that the lookup table is always non-decreasing and peak blade loads across DLC 1.3 are less than the peak loads at 12~ms$^{-1}$.


\section{Power Reference Control}\label{chCon:sec:PR}
\begin{figure}
\centering
\includegraphics[scale=0.85]{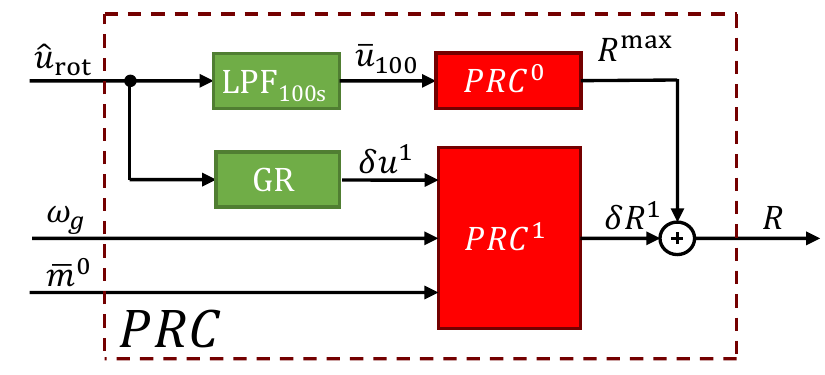}
\caption{The power reference controller, which uses the estimated wind speed $\hat{u}_\text{rot}$, generator speed $\omega_g$, and filtered collective blade load $\bar{m}^0$ as inputs; the output is the power reference factor $R$. The slow power reference control ($PRC^0$) uses a low-pass filtered wind speed ($\bar{u}_{100}$) to determine the maximum power reference $R_\text{max}$, which is decreased (by $\delta R^1$) when problematic transients are expected, estimated using the gust measure $\delta u^1$.}\label{chCon:fig:GR_top}
\end{figure}

Sections~\ref{chCon:sec:PI}--\ref{chCon:sec:MP_PS} have described control elements that can be found elsewhere in the literature, and we have adapted them to work together with the power reference controller described in the following.
The power reference control ($PRC$, Fig.~\ref{chCon:fig:GR_top}) is designed to reduce the power reference factor $R$ when a critical performance variable (structural loading or generator speed) is predicted to exceed a pre-defined threshold.
Otherwise, $R$ is increased to produce more energy.
Because the wind, turbine, and controller behave differently across wind speeds, we use a filtered wind speed signal to determine the maximum allowable power reference factor $R^\text{max}$ using the slow power reference control ($PRC^0$) described in Section~\ref{chCon:sec:PR0}.
To de-rate the turbine during transient events ($R < R^\text{max}$), we use information about the change in wind speed and turbine measurements to decrease $R$ using the transient power reference control ($PRC^1$), a hybrid control system detailed in Section~\ref{chCon:sec:PR1}.
The stability of the power reference control is discussed in Section~\ref{chCon:sec:PR_anal}, along with an analysis of its steady-state behavior and disturbance rejection properties.
A demonstration is presented in Section~\ref{chCon:sec:demo}.

\subsection{Slow Power Reference Control} \label{chCon:sec:PR0}
\begin{figure}
\centering
\includegraphics[scale=0.7]{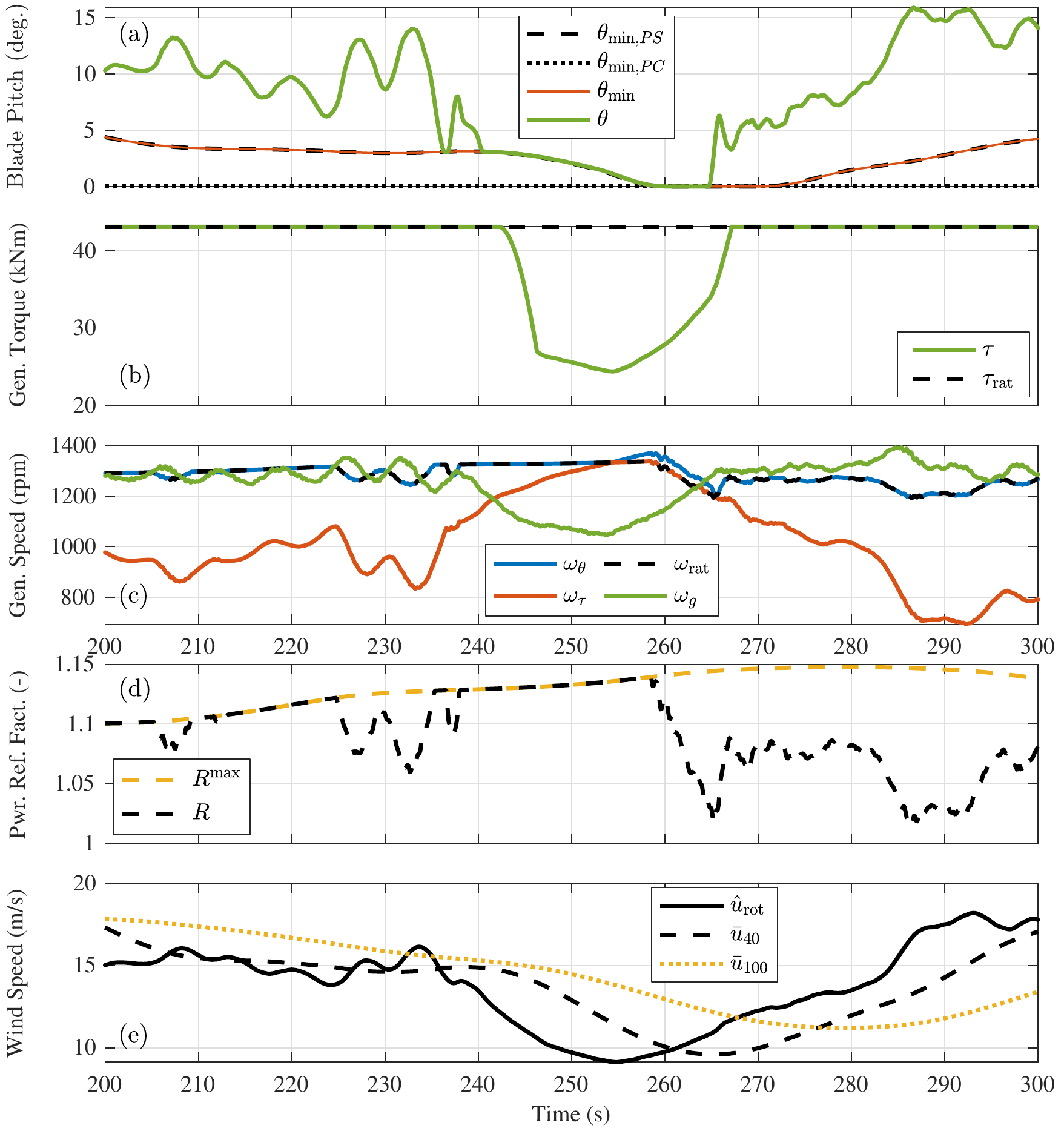}
\caption{
Demonstration of the pitch and torque PI controllers (Section~\ref{chCon:sec:PI} and Fig.~\ref{chCon:fig:PI_control}), set point control (Fig.~\ref{chCon:fig:SSC}), power control (Fig.~\ref{chCon:fig:PC}), minimum pitch peak shaving (Section~\ref{chCon:sec:MP_PS} and Fig.~\ref{chCon:fig:MPPS}(a)), and slow power reference control (Fig.~\ref{chCon:fig:MPPS}(b)).
The generator speed $\omega_g$ is controlled with the blade pitch $\theta$, generator torque $\tau$, and the set points, $\omega_\theta$ and $\omega_\tau$, which depend on the rated generator speed $\omega_\text{rat}$ and set point smoothing controller in Section~\ref{chCon:sec:SSC}.  
The torque and pitch are saturated by $\tau_\text{rat}$ and $\theta_\text{min}$, respectively.
The pitch control generator speed set point $\omega_\theta$ is increased from the rated generator speed $\omega_\text{rat}$ when the torque $\tau$ is less than rated torque $\tau_\text{rat}$ so that the blade pitch is biased towards the minimum pitch setting $\theta_\text{min}$.  
The torque control generator speed set point $\omega_\tau$ decreases when the blade pitch $\theta$ is greater than the minimum pitch setting $\theta_{\text{min},PC}$, which biases the torque control towards its maximum rated torque. 
The rated generator speed $\omega_\text{rat} = R\omega_0$, where $\omega_0$ is the nominal rated generator speed. 
A wind speed estimate $\hat{u}_\text{rot}$ is low-pass filtered ($\bar{u}_{40}$ and $\bar{u}_{100}$) and used as the inputs to lookup tables that determine the minimum pitch setting for peak shaving $\theta_{\text{min},PS}$ and the maximum power reference factor $R^\text{max}$ (an upper bound on $R$).
The minimum pitch limit $\theta_\text{min}$ used by the PI pitch controller is defined in~\eqref{chCon:eq:theta_min}.
Since the power reference factor $R>1$ for this timeseries, the minimum pitch for power control $\theta_{\text{min},PC} = 0$~deg., the optimal pitch angle for this rotor.
}\label{chCon:fig:BaseControl_Demo}
\end{figure}

In the slow power reference controller ($PRC^0$), we control the maximum power reference factor $R^\text{max}$ based on a slow low-pass filtered wind speed $\bar{u}_{100} = \text{LPF}_{100}\big\{\hat{u}_\text{rot}\big\}$, where $\text{LPF}_{\tau_l}$ is defined in~\eqref{chCon:eq:LPF2}.
A time constant of $\tau_l = 100$~sec.~ensures that the maximum power reference factor does not introduce dynamics that would affect the stability of the closed loop system; this is discussed in more detail in Section~\ref{chCon:sec:PR_Stab}.
For the NREL-5MW turbine and the previously described control modules, at wind speeds near cut-out, larger gusts result in greater generator speed transients, so we reduce $R^\text{max}$ at high wind speeds (18--24~ms$^{-1}$) to reduce the average and maximum generator speeds.
Different lookup tables are shown in Fig.~\ref{chCon:fig:MPPS}(b): we place breakpoints 2~ms$^{-1}$ apart and ensure a smooth interpolation using a cubic interpolation.
$R^\text{max}$ (shown in Fig.~\ref{chCon:fig:BaseControl_Demo}(d)) is then the upper bound on the power reference factor $R$ used by the power controller ($PC$) in Section~\ref{chCon:sec:PC}.



Increasing $R^\text{max}$ increases the average power, but also peak generator speeds.
Since increasing $R>1$ results in a lower pitch angle for the same wind speed compared to $R=1$, thrust-based loading is also increased.
The design choice for $R^\text{max}$ versus $\bar{u}_{100}$ (Fig.~\ref{chCon:fig:MPPS}(b)) is ultimately up to the control designer.
Here, we outline the method used to arrive at the lookup tables that produce the results of Section~\ref{chCon:sec:results}.
Since maximum generator speeds depend on the fast power reference control $PRC^1$, presented in Section~\ref{chCon:sec:PR1}, it is important to note that we tune the $PRC^0$ module only after the $PRC^1$ is implemented.
Any changes to $PRC^1$ may require re-tuning $PRC^0$.

After $PRC^1$ is implemented, we perform the following procedure:
\begin{enumerate}
\item Simulate DLC 1.3 and find the worst case generator speed maxima 
\item Increase $R^\text{max}$ across wind speeds until the generator constraint is violated
\item Re-simulate DLC 1.3 and repeat the process, increasing $R^\text{max}$ at wind speeds where there is a gap between the maximum generator speed and the upper bound (see, e.g., the difference between PR-1.100 and PR-1.150 in Fig.~\ref{chCon:fig:MPPS}(b) and the corresponding maximum generator speeds in Fig.~\ref{chCon:fig:ResGen_BL}(b)).  
\item If the constraint is violated, find the $\bar{u}_{100}$ when the maximum occurs and reduce $R^\text{max}$ at the closest breakpoint(s) until the constraint is not violated. 
\end{enumerate}

\subsection{Transient Power Reference Control} \label{chCon:sec:PR1}
In the transient power reference controller $PRC^1$, we predict peaks in the generator speed and blade loads that occur due to wind speed changes, and then de-rate the turbine during these events.
In this article, we use a wind speed estimate $\hat{u}_\text{rot}$, but wind speed measurements could be used to improve the control performance.~\cite{Zalkind2020t}
The other inputs to $PRC^1$ are generator speed $\omega_g$, a known and measurable quantity, and a filtered estimate of the collective blade load component $\bar{m}^0$, which is found by taking the average of the three blade load signals and filtering the signal to eliminate the polluting harmonic frequencies:
\begin{equation}
\bar{m}^0 = \mathcal{F} \left\{ \frac{1}{3} (m_{by,1} + m_{by,2} + m_{by,3}) \right\},
\end{equation}
where $m_{by,i}$ is the flapwise blade load of blade $i$ and $\mathcal{F}$ is a filter designed to eliminate the polluting harmonic from the collective blade load signal.
The filter
\begin{equation}
\mathcal{F} = \text{NF}_\text{3P}(s,\omega_{3\text{P}}) \times \text{LPF}_{1}(s),
\end{equation}
includes a low-pass filter as in~\eqref{chCon:eq:LPF2}, with a time constant of 1~sec., and a moving notch filter
\begin{equation}
\text{NF}(s,\omega_{3\text{P}}) = 
\frac{s^2 + 2 \omega_{3\text{P}} \beta_\text{3P} s + \omega_{3\text{P}}^2}{s^2 + 2 \omega_{3\text{P}} \zeta_\text{3P} s + \omega_{3\text{P}}^2},
\end{equation}
that depends on the changing 3P rotor frequency
\begin{equation}
\omega_{3\text{P}} = \text{LPF}_{100} \left\{ \frac{3 \omega_g}{G} \right\},
\end{equation}
which requires a slow low-pass filtered generator speed for stability and $G=97$ is the gearbox ratio of the NREL-5MW reference turbine.~\cite{Jonkman2009}
The parameters (Table~\ref{chCon:tab:PR1}) of the notch filter ($\zeta_\text{3P},\beta_\text{3P}$) are tuned to reduce the 3P oscillations in the load signal.
\begin{table}[htbp]
  \centering
  \caption{Parameters of the transient power regulator, $PRC^1$, where $\omega_0$ is the nominal rated generator speed, shown in Table~\ref{chCon:tab:PIparams}.}
    \begin{tabular}{rllr}
          & Parameters & Variable & Value \\
    \midrule
    \multicolumn{1}{l}{Blade Load} & Notch Filter Width & $\zeta_\text{3P}$  & 1 \\
    \multicolumn{1}{l}{Filter Parameters} & Notch Filter Depth & $\beta_\text{3P}$  & 0.1 \\
    \midrule
          & Delay Interval & $\Delta t_d$    & 1 s \\
    \multicolumn{1}{l}{Gust Measure} & Number of Delays &$N_d$ & 20 \\
          & Gust Weighting & $w_0$    & 2.5 \\
    \midrule
    \multicolumn{1}{l}{Transient} & Gen. Speed Gain & $d_\omega$ & 40 rpm/(ms$^{-1}$) \\
    \multicolumn{1}{l}{Estimation} & Load Gain & $d_m$    & 750 kNm/(ms$^{-1}$) \\
    \midrule
      & Overspeed Gain & $k_\omega$    & $0.5 / \omega_0$ \\
    \multicolumn{1}{l}{Transient} & Overload Gain & $k_m$    & $3\times 10^{-5}$ (kNm)$^{-1}$ \\
    \multicolumn{1}{l}{De-rating} & Overspeed Limit & $\omega^\text{Lim}$ & 1325 rpm\\
          & Overload Limit & $m^\text{Lim}$ & $9\times 10^3$ kNm \\
    \bottomrule
    \end{tabular}%
  \label{chCon:tab:PR1}%
\end{table}%

\subsubsection{Gust Measure} \label{chCon:sec:gust_rake}
\begin{figure}
\centering
\vspace{3pt}
\includegraphics[width=\linewidth]{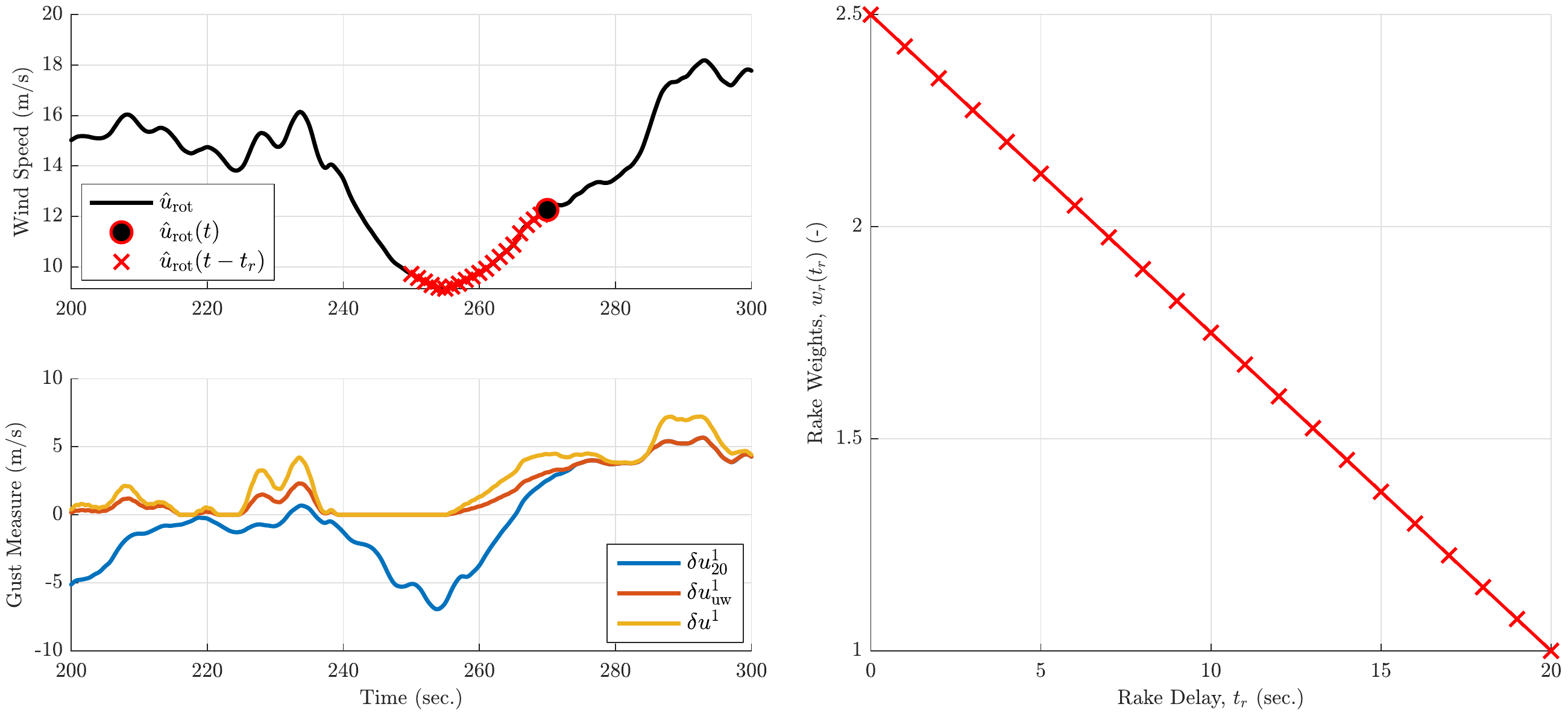}
\caption{The wind speed estimate $\hat{u}_\text{rot}$ sampled at time $t$ and at different delays $t_r$ in the past are used to determine the gust measure (a). Three options are presented (b): a single delay ($\delta u^1_{20}$), with $t_r = \{20\}$, multiple delays ($\delta u^1_\text{uw}$, un-weighted with $w_0 = 1$ and $t_r = \{0,1,2,...,20\}$), and multiple delays with the same $t_r$, but weighted with $w_0 = 2.5$ (c), which is the $\delta u^1$ used to estimate transients for the controller in this article.
}\label{chCon:fig:Gust_Metric}
\end{figure}

Given the generator speed and blade load, we use information about the wind disturbance to estimate transient changes in those signals.
With additional details described in Appendix~\ref{appWSE:chap}, we use an extended Kalman filter to estimate the rotor average wind speed using the generator speed measurement as well as the known blade pitch and generator torque input commands. 
Peak generator speeds and blade loads often occur during lulls in the wind that are followed by increasing wind speeds; we refer to this type of event, shown in Figs.~\ref{chCon:fig:WT_TS} and \ref{chCon:fig:Gust_Metric}(a), as a ``negative gust.''
Our goal is to detect this type of event with a high level of reliability.

First, we sample past wind speeds using multiple delays:
\begin{equation}
U_r = \hat{u}_\text{rot}(t - t_r),
\end{equation}
where $r = \{0,1,2,...,N_d\}$ and the different delays
\begin{equation}
t_r = r \Delta t_d
\end{equation}
are spaced $\Delta t_d$ seconds apart.
The difference between the current wind speed and the set of delayed wind speeds 
\begin{equation}
\Delta U_r = w_r\big[\hat{u}_\text{rot}(t) - \hat{u}_\text{rot}(t-t_r)\big] \label{chCon:eq:dU_set}
\end{equation}
is weighted, giving a greater contribution to sharp wind speed increases:
\begin{equation}
w_r = \frac{1-w_0}{t_{N_d}} t_r + w_0,
\end{equation}
which is shown in Fig.~\ref{chCon:fig:Gust_Metric}(c), where $w_0$ is the weight applied to the zero-delayed wind speed difference.
The maximum $\Delta U_r$
\begin{equation}
\delta u^1 = \max_{r = 0,1,2,...,N_d} \Big\lbrace \Delta U_r \Big\rbrace
\end{equation}
is the gust measure used to predict transients in the generator speed and blade loads.
The process is illustrated in Fig.~\ref{chCon:fig:Gust_Metric} and implemented in Simulink~\cite{Simulink} with a user-defined function and a buffer of the past $N_d \times \Delta t_d$ seconds of the wind speed signal.
When determining the parameters of the gust measure, shown in Table~\ref{chCon:tab:PR1}, the overall goal is to align large values of $\delta u^1$ with peaks in the generator speed.~\cite{Zalkind2020t}

The definition of the gust measure ensures that $\delta u^1 \geq 0$, since the difference between the current wind speed and itself is included in the set $\left\lbrace \Delta U_r \right\rbrace$.
The choice of $N_d$ is a trade off between reducing computational complexity and ensuring that gusts are detected starting from the minimum of a wind speed lull, e.g., near 255~seconds in Fig.~\ref{chCon:fig:Gust_Metric}(a and b).
If only a single delay were used, the increase in wind speed that occurs from 255--265~seconds would not be registered, and the estimated overspeed would be delayed.
%

\subsubsection{Transient Estimation}
We use the gust measure $\delta u^1$ to estimate transients in the generator speed and blade loads:
\begin{eqnarray}
\hat{\omega} &= \omega_g + d_\omega \delta u^1 \label{chCon:eq:w_est} \\
\hat{m} &= \bar{m}^0 + d_m \delta u^1, \label{chCon:eq:m_est}
\end{eqnarray}
where  $d_\omega$ and $d_m$ are the transient gains for generator speed and blade load, respectively.
The transient gains (in Table~\ref{chCon:tab:PR1}) are tuned by analyzing the step response to a system only controlled by the PI torque and pitch controllers.~\cite{Zalkind2020t}

More sophisticated methods for estimating transients surely exist, e.g., using model predictive control.
However, these methods rely on using accurate models, require more computational effort, and make analyzing their behavior more difficult.
The goal of this article is to provide a proof-of-concept demonstration for safely increasing the power output using power reference control.

\subsubsection{De-rating using a hybrid automata} \label{chCon:sec:fDR_SM}
\begin{figure}
\centering
\includegraphics[scale=0.85]{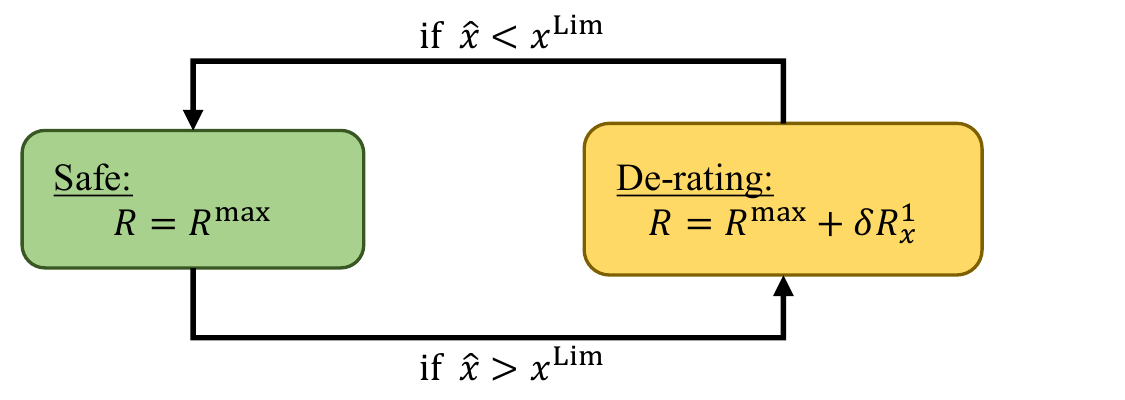}
\caption{State machine for the de-rating signal.
}\label{chCon:fig:stateMachine}
\end{figure}

When the estimated transients $\hat{\omega}$ or $\hat{m}$ exceed thresholds $\omega^\text{Lim}$ and $m^\text{Lim}$, respectively, the transient power reference control $PRC^1$ reduces the power reference factor so that $R < R^\text{max}$.
The decrement amounts
\begin{equation}
\delta R^1_x = - k_x\hat{x}, \quad \text{where}\; x= \{\omega,m\} \label{chCon:eq:propDerate}
\end{equation}
are computed simultaneously for both the generator speed and blade load.
The power reference factor is then computed as
\begin{equation}
R = R^\text{max}(\bar{u}_{100}) + \text{min}\{\delta R_\omega^1,\delta R_m^1\}, \label{chCon:eq:R}
\end{equation}
where $R^\text{max}(\bar{u}_{100})$ is given in Fig.~\ref{chCon:fig:MPPS}(b) for several cases and $R$ is the input to the power controller ($PC$) in Section~\ref{chCon:sec:PC}.
When the estimated states are less than the thresholds, the controller acts as it normally would, with $R = R^\text{max}$.  
We can implement and model the controller with a hybrid automata (Fig.~\ref{chCon:fig:stateMachine}) and two control states: safe and de-rating.

\subsection{Power Reference Control Analysis} \label{chCon:sec:PR_anal}
For analysis purposes, each state in the hybrid system can be modeled separately, while we seek to answer the following questions about the $PRC$:
\begin{itemize}
\item \textbf{Stability:} Under what conditions are each of the states (safe and de-rating) stable? Is the transition between the states stable?
\item \textbf{Disturbance Rejection:} Given the system parameters, can we predict the maximum transients that will occur when using the power regulator?
\item \textbf{Steady State:} Does the system return to the safe state after a problematic transient event? 
\end{itemize}


Until this point, we have presented the non-linear control system that is used in DLC simulations.
To answer the above questions, we analyze linear models that represent the two states (safe and de-rating) and include the rotor dynamics, pitch actuator, and PI pitch control system, with a wind disturbance ($u_\text{rot}$) and generator speed reference ($\omega_\theta$) inputs.
We ignore the nonlinear aspects of the set point controller in Section~\ref{chCon:sec:SSC} and assume $\omega_\theta = \omega_\text{rat}$.
We then compare properties found in the hybrid linear system with nonlinear simulation results to verify the closed-loop behavior of the system~\cite{Zalkind2020t}; the results are summarized next.

\subsubsection{Stability Analysis} \label{chCon:sec:PR_Stab}

\paragraph{Stability of Safe State}
In the safe state, the reference $R= R^\text{max}(\bar{u}_{100})$ varies slowly, since $\bar{u}_{100}$ is a filtered wind speed with a time constant of 100~s.
Thus, we assume that $\omega_\theta = R \omega_0$ is constant and we determine the stability margins of the underlying system, without reference control.
The linearized system suggests that there is an upper bound on $k_{P,\theta}$ and $k_{I,\theta}$ for the state matrix to be Hurwitz.~\cite{Zalkind2020t}
If we use a gain factor $G_\text{fact}$ to increase both PI gains from their originally designed value in Table~\ref{chCon:tab:PIparams}, such that
\begin{equation}
k_{P,\theta}^\text{new} = G_\text{fact} k_{P,\theta} \quad \text{and} \quad
k_{I,\theta}^\text{new} = G_\text{fact} k_{I,\theta},
\end{equation}
we find some $G_\text{fact} > 1$ that makes the linear system unstable,~\cite{Zalkind2020t} similar to a gain margin without considering the theoretical, frequency-domain specifics.
Instability is determined in nonlinear simulations when there are large pitch variations that saturate rate limiters; technically, the nonlinear systems is stable, but it exhibits undesirable oscillatory pitch control.

%
%
%
%

\paragraph{Stability of the De-rating State}
When $\hat{\omega} > \omega^\text{Lim}$, the turbine is de-rated according to~\eqref{chCon:eq:R}.
Thus, the change in power reference $\delta R^1_\omega = -k_\omega \hat{\omega}$ is proportional to the generator speed transient estimation~\eqref{chCon:eq:w_est} and the linearized change in generator speed reference is
\begin{equation}
\delta\omega_\theta = -k_\omega \omega_0 \omega_g - k_\omega \omega_0 d_\omega \delta u^1. \label{chCon:eq:PR1_control}
\end{equation}
The second term in~\eqref{chCon:eq:PR1_control} functions like a feedforward input and the first term alters the state dynamics due to its dependence on $\omega_g$.
When the system is in the de-rating state, the state dynamics are as if the PI gains are increased by a factor of $(1+k_\omega \omega_0)$, namely
\begin{equation}
k_{P,\theta} \leftarrow  k_{P,\theta}(1+k_\omega \omega_0)\quad \text{and} \quad
k_{I,\theta} \leftarrow  k_{I,\theta}(1+k_\omega\omega_0).
\end{equation}
and the gain factor $G_\text{fact}$ that causes the underlying system to be unstable provides an upper bound constraint for stability on the overspeed gain $k_\omega$:
\begin{equation}
k_\omega \omega_0 + 1 < G_\text{fact}.  \label{chCon:eq:derateStab}
\end{equation}
De-rating due to load ($\hat{m} > m^\text{Lim}$) results in similar, but less intuitive, state dynamics and an upper bound on $k_m$.



\paragraph{Stability of Transition}
Using the design parameters in Table~\ref{chCon:tab:PR1}, the linear system that represents each hybrid state is stable; thus, there exists a valid Lyapunov function for each that decreases in time.
Furthermore, we can find a common Lyapunov function using a linear matrix inequality that is positive definite with a derivative that is negative definite, which implies that the states of the hybrid system are always decreasing, regardless of any transitions.
The common Lyapunov function can be used as a measure of how close the states converge to an equilibrium point.
Using nonlinear simulations, we observe that values of  $G_\text{fact}$ and $k_\omega$ that cause unstable linear systems result in nonlinear states that do not converge nearly as close to the equilibrium points as stable parameters.~\cite{Zalkind2020t}

\subsubsection{Disturbance Rejection}
When operating in the de-rating state~\eqref{chCon:eq:PR1_control},
the first term has the effect of increasing the PI gains, which increases the bandwidth of the closed-loop system and reduces the peak transient in generator speed during a disturbance input.
The second term functions as a ``feedforward,'' where the generator speed reference $\omega_\theta$ is reduced in proportion to the estimated disturbance $\delta u^1$ and induces an offset in $\omega_g$.
Both effects, the increase in bandwidth and reduction in reference, act to reduce the peak transients when $k_\omega$ is increased.

Simulation results also suggest that maximum generator speeds decrease with an increasing $k_\omega$, but only up until some point; there are several reasons why this may occur.
There is a delay between the actual wind speed disturbance and wind speed estimate (WSE), which propagates to the estimated generator speed transients and de-rating of the turbine, reducing performance in terms of peak generator speed reduction.
Additionally, increasing $k_\omega$ by too much reduces the stability margin and leads to poor performance in general; extra pitch actuation increases loads and generator speed variation.
Generally, it is difficult to control the response to gusts consistently because varied turbine states and random wind speed disturbances make controlling peak generator speeds with a high level of certainty difficult in a realistic operating environment.

\subsubsection{Steady-State Analysis}
When the controller is in the de-rating state, we want the system to return to the safe state when the problematic wind disturbance has passed, so that the greatest amount of energy is captured.
Linear models suggest that the steady-state behavior returns the system to the safe state.~\cite{Zalkind2020t}
Here, we present a logical argument for the system returning to the safe state.

Because the PI pitch (and torque) controllers include integral control, $\omega_g \rightarrow \omega_\theta$ eventually.
When the system is in the de-rating state
\begin{equation}
\omega_\theta < \omega_0 R^\text{max} < \omega^\text{Lim}. \label{chCon:eq:SS_Setpoints}
\end{equation}
Thus $\omega_g < \omega^\text{Lim}$ eventually and the system returns to the safe state if~\eqref{chCon:eq:SS_Setpoints} is true.
If the designer chooses an $R_\text{max}$ such that~\eqref{chCon:eq:SS_Setpoints} is not true and $\omega^\text{Lim} < \omega_0 R^\text{max}$, then the system will remain in the de-rating state and diminishing benefits in energy capture will occur.

\subsection{Demonstration} \label{chCon:sec:demo}
Throughout this article, we referred to Fig.~\ref{chCon:fig:WT_TS}, where a problematic gust event occurs during extreme turbulence with a mean wind speed of 18~ms$^{-1}$.
The controller transitions between above- and below-rated operation during this event and the power drops to nearly half of the rated value.
When the wind speed then increases, blade and tower loads peak, as does the generator speed.

\begin{figure}
\centering
\vspace{3pt}
\includegraphics[scale=0.75]{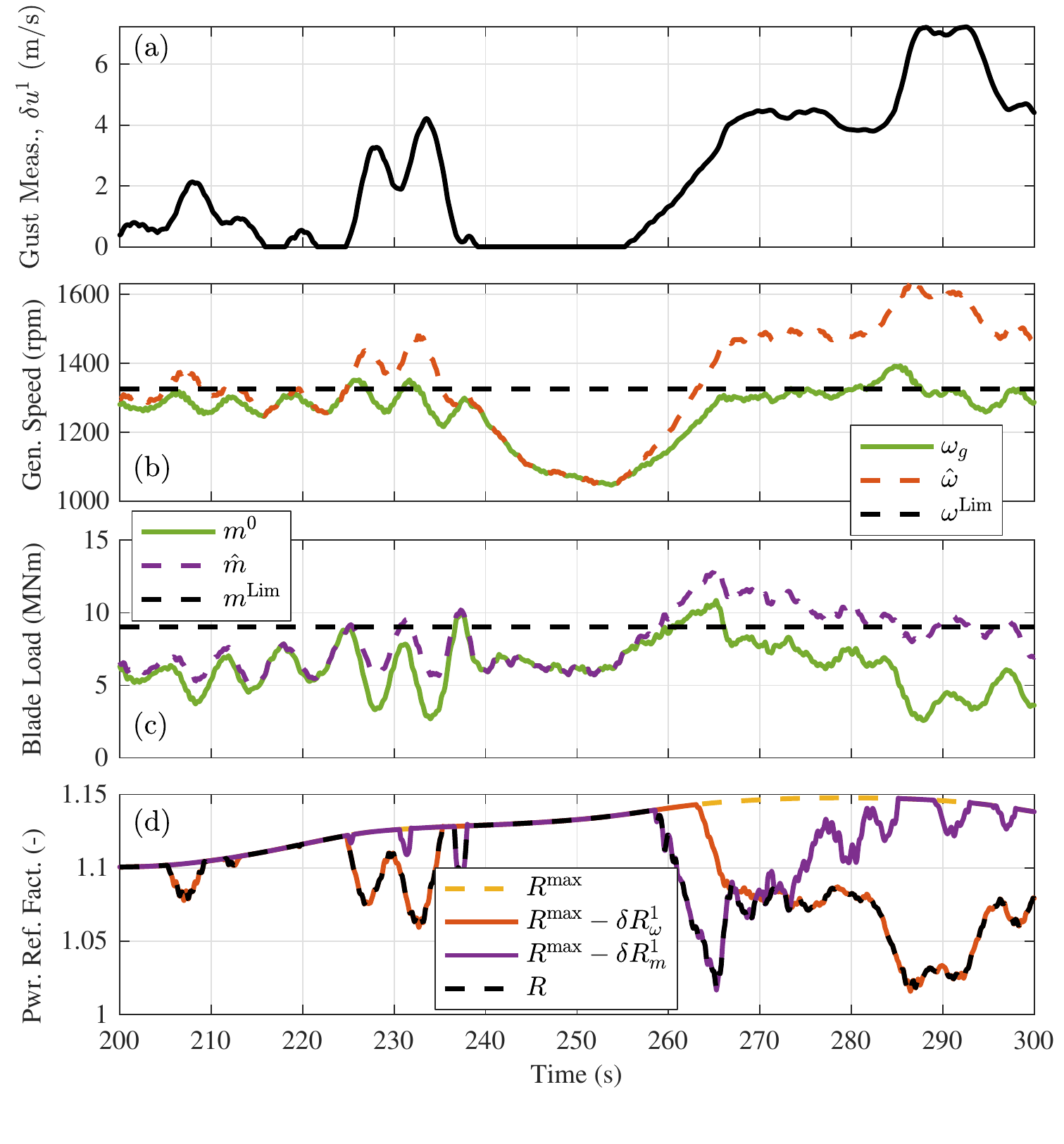}
\caption{Demonstration of $PRC^1$. The gust measure $\delta u^1$ leads to transient estimates of the generator speed $\hat{\omega}$ and collective blade load $\hat{m}$.  When the transient estimates exceed the limits $\omega^\text{Lim}$ or $m^\text{Lim}$, the transient derating signals $\delta R_\omega^1$ and $\delta R_m^1$ reduce the power reference factor $R$ from the maximum reference factor $R^\text{max}$; these signals, also shown in Fig.~\ref{chCon:fig:BaseControl_Demo}(d), are used by the power controller.}\label{chCon:fig:PR1_demo}
\end{figure}

The $PRC$ determines the gust measure ($\delta u^1$ in Fig.~\ref{chCon:fig:Gust_Metric}), which is used to estimate the generator speed and blade load transients and de-rate the turbine (Fig.~\ref{chCon:fig:PR1_demo}).
The maximum power reference $R^\text{max}$ is determined using the slow low-pass filtered wind speed ($\bar{u}_{100}$) as shown in Fig.~\ref{chCon:fig:BaseControl_Demo}(e) along with the minimum pitch limit for peak shaving.

The power controller uses the power reference $R$ to determine the rated generator speed $\omega_\text{rat}$ and minimum pitch setting for power control $\theta_{\text{min},PC}$, as shown in Fig.~\ref{chCon:fig:BaseControl_Demo}(a, c, and d).
The set point smoothing controller determines the set points for the torque and pitch controllers, $\omega_\tau$ and $\omega_\theta$, respectively, which regulate the generator speed $\omega_g$ using the torque $\tau$ and pitch $\theta$ inputs to the turbine (Fig.~\ref{chCon:fig:BaseControl_Demo}a, b, and c).
Besides controlling the generator speed, the torque and pitch inputs determine the generator power $P$, shown in Fig.~\ref{chCon:fig:WT_TS}, along with the blade load $m_{by,1}$ and rotor thrust $T$.
Several similar events like the one in this demonstration occur when simulating a full set of design load cases, which we report on in Section~\ref{chCon:sec:results}.

\section{Simulation Results} \label{chCon:sec:results}
Next, we compare controllers using modules common in practice (described in Sections~\ref{chCon:sec:PI}--\ref{chCon:sec:MP_PS}) with controllers that use the power reference control described in Section~\ref{chCon:sec:PR}.
The simulation results, from design load case (DLC) simulations, are discussed in terms of energy capture, maximum generator speeds, and structural loading.

First, we present a set of baseline controllers:
\begin{itemize}
\item \textbf{NREL-5MW}: the reference controller~\cite{Jonkman2009} commonly used as a benchmark for comparison in the wind turbine community
\item \textbf{BL-1.000}: our baseline for comparison, which includes the PI torque and pitch controllers (Section~\ref{chCon:sec:PI}), set point smoothing controller (Section~\ref{chCon:sec:SSC}), and the power controller (Section~\ref{chCon:sec:PC}) with a constant power reference $R=1$.
\item \textbf{BL-1.050}: our ``boosted'' baseline, which is the same as the BL-1.000, but with an increased PI pitch control bandwidth ($\omega_{\text{reg},\theta} = 0.4$~rad~s$^{-1}$ versus 0.275~rad~s$^{-1}$), minimum pitch peak shaving (Section~\ref{chCon:sec:MP_PS}), and only the slow power reference controller $PRC^0$ (Section~\ref{chCon:sec:PR0}) using a variable $R^\text{max}$ up to 1.050, shown in Fig.~\ref{chCon:fig:MPPS}(b). A wind speed estimate $\hat{u}_\text{rot}$ is used for the wind input.
\end{itemize}

The performance of these baseline controllers are compared with power reference (PR) controllers that make full use of the power reference control in Section~\ref{chCon:sec:PR}:
\begin{itemize}
\item \textbf{PR-1.100}: a controller with all of the previously described control modules, including the transient power reference control ($PRC^1$, Section~\ref{chCon:sec:PR1}) and a constant $R^\text{max} = 1.100$, using a wind speed estimate $\hat{u}_\text{rot}$ for the wind signal and a PI pitch control bandwidth of $\omega_{\text{reg},\theta} = 0.275$~rad s$^{-1}$.
\item \textbf{PR-1.150}: the same as PR-1.100, but with an additional power boost below 18~m s$^{-1}$, up to $R^\text{max} = 1.150$.
\end{itemize}

\begin{figure}
\centering
\includegraphics[width=0.475\linewidth]{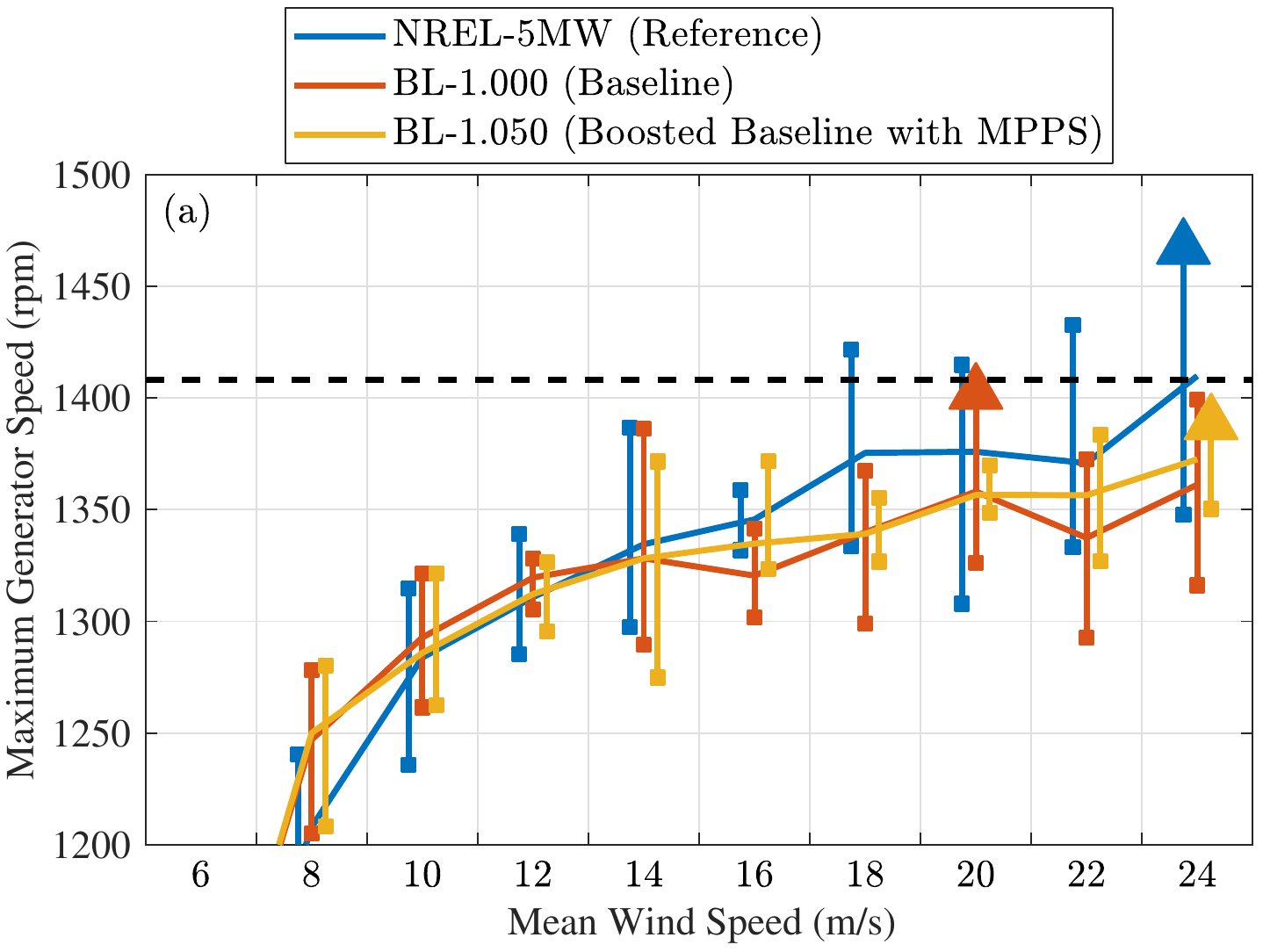}
\hfill
\includegraphics[width=0.475\linewidth]{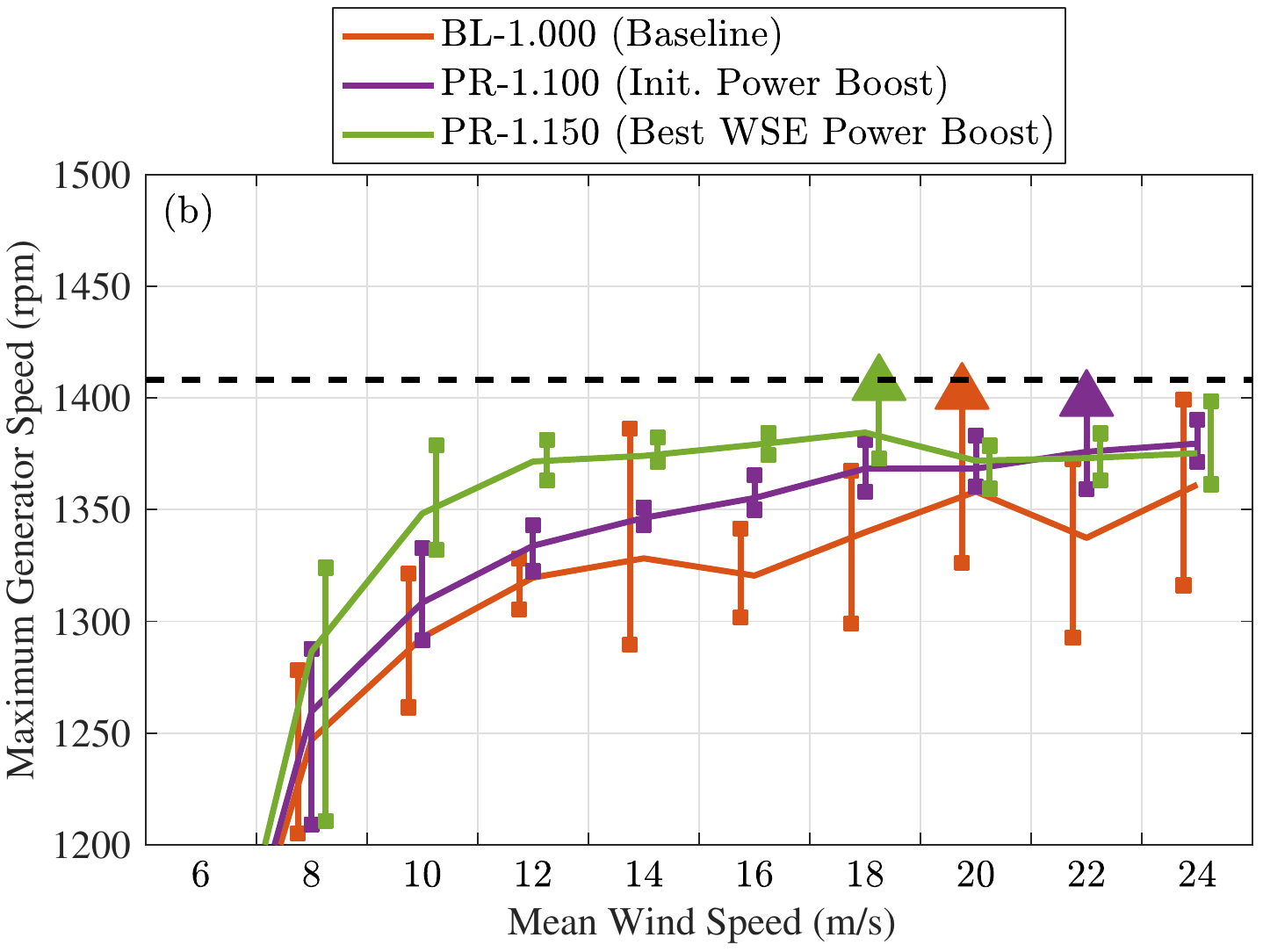}\\
\vspace{0.5cm}
\includegraphics[width=0.475\linewidth]{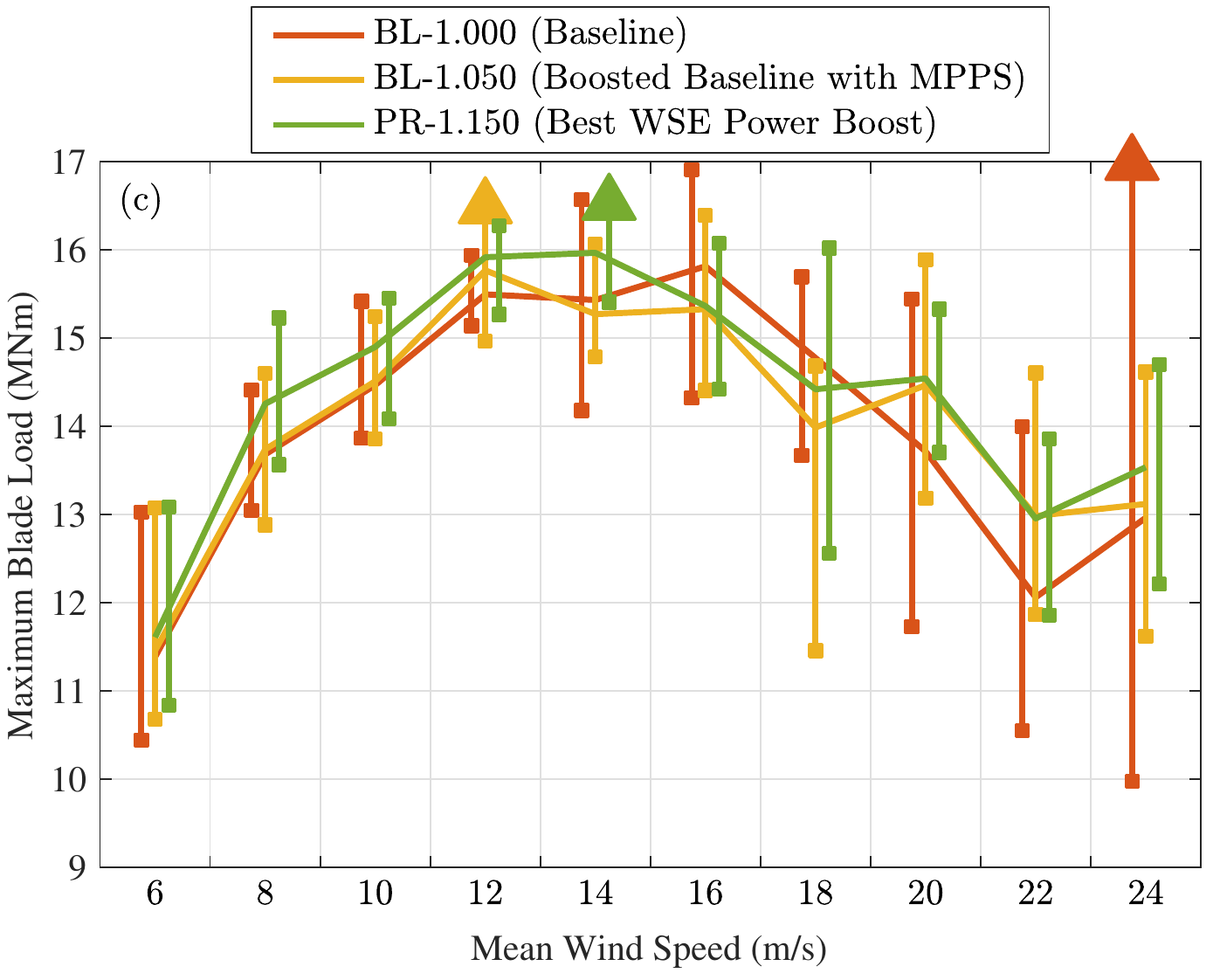}
\hfill
\includegraphics[width=0.475\linewidth]{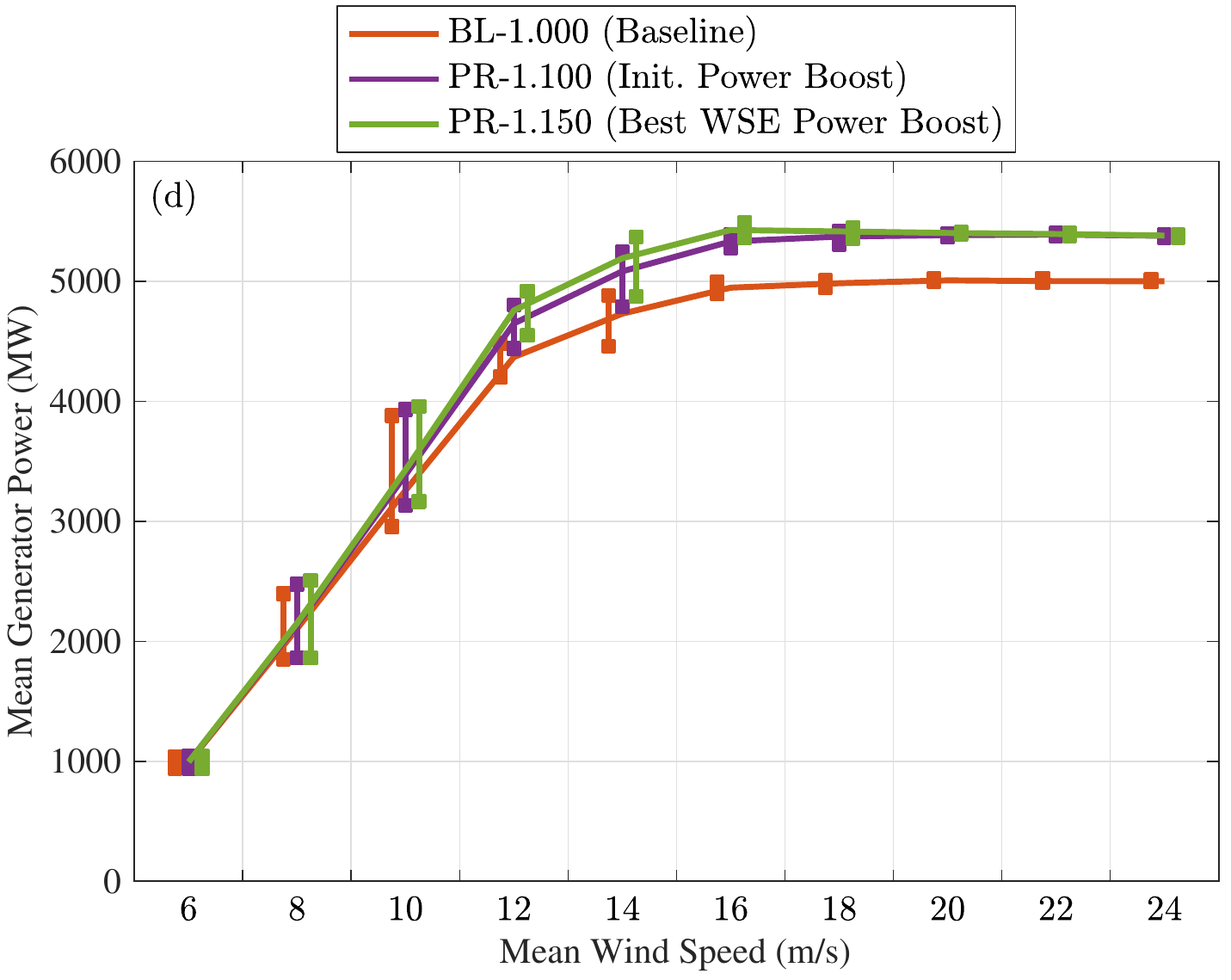}
\caption{Maximum generator speeds (a,b), maximum blade loads (c), and average generator power (d) for the various controllers described in Section~\ref{chCon:sec:results}.  At each mean wind speed, the maximum generator speeds, maximum blade loads, and mean generator power of six turbulent wind simulations span the vertical lines between the squares ($\blacksquare$); the overall maxima is denoted with a triangle ($\blacktriangle$).  Note that the vertical lines are offset horizontally for comparison purposes.} \label{chCon:fig:ResGen_BL}
\end{figure}

\begin{table}[htbp]
  \centering
  \caption{Summary of results for various controllers, detailing the energy capture, blade loads, rotor thrust, maximum generator speed, and tower base fore-aft damage equivalent loading (with each measure compared to the BL-1.000 controller). The maximum allowable generator speed limit is 120\% of the nominal rated generator speed, or 1408 rpm.}
\begin{tabular}{lcccc}
          & \multicolumn{2}{c}{Lifetime Average Power (MW)} & \multicolumn{2}{c}{Blade Load (MNm)} \\
    Controller & DLC 1.2 & DLC 1.3 & Characteristic & Maximum \\
    \midrule
    NREL-5MW Ref. & 1834.6 (+0.61\%) & 2235.3 (+0.66\%) & 16.67 (+5.37\%) & 17.92 (+5.60\%) \\
    BL-1.000 & 1823.4 (-) & 2220.6 (-) & 15.82 (-) & 16.97 (-) \\
    BL-1.050 & 1837.6 (+0.78\%) & 2242.5 (+1.00\%) & 15.77 (-0.32\%) & 16.48 (-2.89\%) \\
    \midrule
    PR-1.100 & 1901.8 (+4.30\%) & 2328.0 (+4.84\%) & 15.55 (-1.71\%) & 16.59 (-2.24\%) \\
    PR-1.150 & 1925.4 (+5.60\%) & 2361.2 (+6.33\%) & 15.97 (+0.95\%) & 16.52 (-2.65\%) \\
    \midrule
          &       &       &       &  \\
          & \multicolumn{2}{c}{Rotor Thrust (MN)} & Max. Gen. & Tower Base Fore- \\
    Controller & Characteristic & Maximum & Speed (rpm) & Aft DEL (MNm) \\
    \midrule
    NREL-5MW Ref. & 1.02 (+4.45\%) & 1.19 (+7.09\%) & 1467  & 20.8 (-9.17\%) \\
    BL-1.000 & 0.977 (-) & 1.11 (-) & 1402  & 22.9 (-) \\
    BL-1.050 & 0.978 (+0.54\%) & 1.04 (-6.60\%) & 1384  & 25.6 (+11.8\%) \\
    \midrule
    PR-1.100 & 0.909 (-7.00\%) & 0.98 (-11.6\%) & 1399  & 21.5 (-6.11\%) \\
    PR-1.150 & 0.964 (-1.31\%) & 1.05 (-5.17\%) & 1406  & 23 (+0.43\%) \\
    \bottomrule
    \end{tabular}%
  \label{chCon:tab:results}%
\end{table}%
To measure the performance of the various controllers, we simulate each using the same set of wind fields defined in DLCs 1.2 and 1.3 (NTM and ETM, respectively). 
The full set of results are summarized in Table~\ref{chCon:tab:results} in terms of the performance measures described in Section~\ref{chCon:sec:metrics}.
Each controller is compared in terms of its energy capture (or lifetime average power) in both DLCs 1.2 and 1.3; an example for DLC~1.3 is shown in Fig.~\ref{chCon:fig:ResGen_BL}(d).
The controllers are designed using a maximum generator speed constraint of 1408~rpm.
In Fig.~\ref{chCon:fig:ResGen_BL}(a and b), the maximum generator speeds $\omega_g^\text{max}$ are shown for each of the controllers.


The maximum blade loads are depicted in Fig.~\ref{chCon:fig:ResGen_BL}(c), while the maximum and characteristic rotor thrust and blade loads are shown in Table~\ref{chCon:tab:results}.
DLC 1.3 (ETM) is used to determine the generator speed, load, and thrust maxima and characteristic loading.
DLC 1.2 is used to determine the tower base fore-aft lifetime damage equivalent loading (DEL) for the various controllers.
The fore-aft DEL is also a good measure of extra pitch actuation, the primary cost of using the power regulator.

\subsection{Discussion of Results}
The NREL-5MW reference controller results in a generator speed constraint violation during DLC 1.3 simulations ($\omega_g^\text{max}=1467$~rpm at 24~ms$^{-1}$, which exceeds the upper limit of 1408~rpm).
Our baseline controller, BL-1.000, was tuned so that $\omega_g^\text{max} < 1408$~rpm for all DLC 1.3 simulations and is used for comparison with all other controllers.

If we increase the bandwidth of the PI pitch controller via $\omega_\text{reg}$ (from 0.275~rad s$^{-1}$ to 0.4~rad s$^{-1}$), as we do for the BL-1.050, the generator speed has less variation (and overshoot) from its set point and we are able to boost the power reference up to $R^\text{max} = 1.050$.
Based on our stability analysis (Section~\ref{chCon:sec:PR_Stab}), the increased PI pitch gains reduce the stability margin; thus, we could not use the transient power reference control with the BL-1.050 controller.
To decrease blade loads (especially at high wind speeds), we must include the minimum pitch control; notice, in Fig.~\ref{chCon:fig:ResGen_BL}(c), that the maximum blade load for the baseline (BL-1.000) control occurs near cut out.
The steady power regulator must also be used to decrease $R^\text{max}$ near rated wind speeds.
Otherwise, an ETM simulation at 14~m s$^{-1}$ has a generator speed peak that exceeds 1408~rpm.
Compared with the BL-1.000 controller, the BL-1.050 increases the lifetime average power by 1\% and the tower base fore-aft DEL by 12\%, while there is only a small change in the other performance measures.

The transient power reference control $PRC^1$ (with the original PI pitch controller using $\omega_\text{reg} = 0.275$~rad s$^{-1}$) allows us to increase $R^\text{max}$ because of the reduced spread in $\omega_g^\text{max}$, compared to the baselines, as shown in Fig.~\ref{chCon:fig:ResGen_BL}(a and b).
PR-1.100 increases the lifetime average power by 4\% compared to the baseline, but
there is a gap between the maximum generator speeds below 18~ms$^{-1}$ and the hard upper bound, which implies that $R^\text{max}$ can be increased even more at these wind speeds.
Thus, the slow power reference control $PRC^0$ can be used to increase $R^\text{max}$ more below 18~ms$^{-1}$, resulting in the largest amount of power boost achieved (using a WSE and other control modules described in this article) that satisfies all the constraints. 
Compared to the BL-1.000, we realize an increase in lifetime average power of 5.6\% during DLC 1.2 and 6.3\% in DLC 1.3.

\section{Conclusions}\label{sec:conclusion}
In this article, we present a controller that increases energy capture while maintaining the same generator speed and blade load limits.
Because annual energy production has a direct impact on the levelized cost of energy, while loads and generator speed behave more like constraints on the overall design, we believe these goals more accurately reflect system-level turbine design goals.

We achieve these goals by increasing the power reference and then de-rating (or decreasing) the power reference only when a critical performance variable would otherwise exceed some threshold.
In this article, we control peak transients in the generator speed and flapwise blade loads, though the method could be applied more generally to other loads or values.
This power reference control was implemented and analyzed using a hybrid controller, which interacts with a series of modular control elements, reflecting collaborative design practices and allowing control engineers to utilize the various modules we have presented.
The power reference controller provides an input to a power controller, which reduces generator speed and blade load transients while increasing power output.
Peak load shaving is implemented using a minimum pitch limit.
A set point controller determines the generator speed set points to provide minimal interaction between the PI pitch and torque controllers, which supply the direct inputs to the turbine.

Using the power reference controller with a wind speed estimate increases lifetime average power by 5--6\% compared to a baseline controller with a constant power reference.
Generator speeds stay below a pre-defined threshold, peak blade loads are reduced, and tower fatigue increases by less than 1\%.
By reconsidering our control goals and aligning them with the system-level goals of wind turbine design, we can improve performance from a system-level perspective.

\subsection*{Acknowledgements}
The information, data, and work presented herein was funded in part by the Advanced Research Projects Agency - Energy (ARPA-E), U.S. Department of Energy, under Award Number DE-AR0000667. 
Support from a Palmer Endowed Chair Professorship is also gratefully acknowledged.
The views and opinions of the authors expressed herein do not necessarily state or reflect those of the United States Government or any agency thereof.

\appendix

\section{Wind Speed Estimator} \label{appWSE:chap}
The output of a wind speed estimator (WSE) is used in multiple control modules described in this article:
\begin{itemize}
\item In Section~\ref{chCon:sec:MP_PS}, a low-pass filtered wind speed estimate is used to determine the lower limit of the pitch controller during near- and above-rated operation for peak load reduction,
\item In Section~\ref{chCon:sec:PR0}, a filtered wind speed estimate is used to change the maximum allowable power reference, and
\item In Section~\ref{chCon:sec:PR1}, the wind speed estimate is translated into a gust measure that is used to estimate problematic transients and de-rate the turbine from the maximum allowable power reference.
\end{itemize}

Using known inputs and outputs of the turbine, the WSE provides an estimated wind speed signal.
The WSE in this article is implemented as a discrete extended Kalman filter (EKF), with state dynamics for the drivetrain and mean and turbulent wind speed dynamics. 
The drivetrain degree-of-freedom is primarily driven by the power coefficient, which is estimated from sampled simulation data.~\cite{Zalkind2020t}
The wind speed dynamics include a slowly varying mean wind speed and a quickly varying turbulent component, which combine to provide the wind speed estimate.~\cite{Knudsen2013}
A discrete EKF provides a simple implementation; similar examples~\cite{Simley2016,BarShalom2003} and additional details of the WSE used in this article are provided in previous work.~\cite{Zalkind2020t}

An example wind speed estimate is shown in Fig.~\ref{chCon:fig:Gust_Metric}, which is used to estimate transients in Fig.~\ref{chCon:fig:PR1_demo}, and also set the maximum power reference in Fig.~\ref{chCon:fig:BaseControl_Demo}(d) and minimum pitch angle in Fig.~\ref{chCon:fig:BaseControl_Demo}(a).
Overall performance measures of the WSE used in this article are provided in previous work.~\cite{Zalkind2020t}
The goal was to design a WSE with low bias, for accurately determining the operating wind speed, and one with adequate disturbance estimation, measured via the relative degree of explanation (RDE).~\cite{Soltani2013}
Our WSE has RDE values greater than 75\% across the operating wind speeds, which provides the controller with adequate disturbance estimation; it is not necessarily the best WSE possible, but was used to attain the results in this article.

\bibliography{load}

\begin{thebibliography}{10}
\providecommand \doibase [0]{http://dx.doi.org/}%

\bibitem{Stehly2018}
Stehly T, Beiter P, Heimiller D, Scott G. {2017 cost of wind energy review}.
  Tech. Rep. NREL/TP-5000-63267, National Renewable Energy Laboratory;
  \url{https://www.nrel.gov/docs/fy18osti/72167.pdf}:   2018.

\bibitem{Zalkind2019c}
Zalkind DS, Ananda GK, Chetan M, Martin DP, Bay CJ, Johnson KE, Loth E,
  Griffith DT, Selig MS, Pao LY. System-level design studies for large rotors.
  {\it Wind Energy Science} 2019\string; 4(4)\string: 595--618.
\newblock \href {\doibase 10.5194/wes-4-595-2019} {doi: 10.5194/wes-4-595-2019}

\bibitem{PAO2021}
Pao LY, Zalkind DS, Griffith DT, Chetan M, Selig MS, Ananda GK, Bay CJ, Stehly
  T, Loth E. Control co-design of 13 MW downwind two-bladed rotors to achieve
  25
\newblock \href {\doibase https://doi.org/10.1016/j.arcontrol.2021.02.001}
  {doi: https://doi.org/10.1016/j.arcontrol.2021.02.001}

\bibitem{Bortolotti2016}
Bortolotti P, Bottasso CL, Croce A. {Combined preliminary–detailed design of
  wind turbines}. {\it Wind Energy Science} 2016\string; 1(1)\string: 71--88.
\newblock \href {\doibase 10.5194/wes-1-71-2016} {doi: 10.5194/wes-1-71-2016}

\bibitem{Hansen2013}
Hansen MH, Henriksen LC. {Basic DTU wind energy controller}. Tech. Rep. E-0018,
  Technical University of Denmark;
  \url{https://orbit.dtu.dk/files/56263924/DTU_Wind_Energy_E_0028.pdf}:   2013.

\bibitem{Aho2012}
{Aho} J, {Buckspan} A, {Laks} J, {Fleming} P, {Jeong} Y, {Dunne} F,
  {Churchfield} M, {Pao} L, {Johnson} K. A tutorial of wind turbine control for
  supporting grid frequency through active power control. In:  {\it Proc. American Control Conference}; 2012\string: 3120-3131

\bibitem{SWP}
{Siemens Wind Power} . {Siemens power boost function}. {online};  2014.
\newblock
  \url{http://www.energy.siemens.com/us/pool/hq/services/renewable-energy/wind-power/swp-power-boost-function.pdf}.

\bibitem{VWS2014}
{Vestas Wind Systems} . {Vestas launches new upgrades to increase output of
  installed turbines}. {online};  2014.
\newblock
  \url{https://www.vestas.com/en/media/~/media/dd6c48580743401b8710fa22cc9c68e5.ashx}.

\bibitem{Petrovic2017}
Petrovi{\'{c}} V, Bottasso CL. Wind Turbine Envelope Protection Control Over
  the Full Wind Speed Range. {\it Renewable Energy} 2017\string; 111\string:
  836--848.
\newblock \href {\doibase 10.1016/j.renene.2017.04.021} {doi:
  10.1016/j.renene.2017.04.021}

\bibitem{Kanev2017}
Kanev S. {Extreme turbulence control for wind turbines}. {\it Wind Engineering}
  2017.
\newblock \href {\doibase 10.1177/0309524X17723204} {doi:
  10.1177/0309524X17723204}

\bibitem{Zalkind2019b}
{Zalkind} DS, {Pao} LY. Constrained wind turbine power control. In:  {\it Proc.~American Control Conference}; 2019\string: 3494-3499

\bibitem{Zalkind2020t}
Zalkind DS. {\it Methods for enabling control collaboration during wind turbine design}. PhD thesis. University of Colorado Boulder, Boulder, CO, USA;  2020.

\bibitem{Commission2005}
{International Electrotechnical Commission} . {Wind turbines - part 1: design
  requirements}. Tech. Rep. IEC 61400-1:2005(E), ; :   2005.

\bibitem{Griffith2014}
Griffith DT, Richards PW. {The SNL100-03 blade: design studies with flatback
  airfoils for the Sandia 100-meter blade}. Tech. Rep. SAND2014-18129, Sandia
  National Laboratory;
  \url{http://energy.sandia.gov/wp-content/gallery/uploads/dlm_uploads/1418129.pdf}:
  2014.

\bibitem{Ning2014}
Ning A, Damiani R, Moriarty PJ. {Objectives and constraints for wind turbine
  optimization}. {\it Journal of Solar Energy Engineering} 2014\string;
  136(4)\string: 041010.
\newblock \href {\doibase 10.1115/1.4027693} {doi: 10.1115/1.4027693}

\bibitem{Hayman2012}
Hayman GJ. {MLife theory manual for version 1.00}. Tech. Rep. NREL/TP-XXXXX,
  National Renewable Energy Laboratory;
  \url{https://nwtc.nrel.gov/system/files/MLife_Theory.pdf}:   2012.

\bibitem{Jonkman2009}
Jonkman J, Butterfield S, Musial W, Scott G. {Definition of a 5-MW reference
  wind turbine for offshore system development}. Tech. Rep. NREL/TP-500-38060,
  National Renewable Energy Laboratory;
  \url{https://www.nrel.gov/docs/fy09osti/38060.pdf}:   2009.

\bibitem{OpenFAST}
NREL . OpenFAST. Version 2.2.0. Online;  2019.
\newblock \url{https://github.com/OpenFAST/openfast}.

\bibitem{Jonkman2012}
Jonkman B, Kilcher L. {TurbSim user's guide: version 1.06.00}. Tech. Rep.
  TP-500-39797, National Renewable Energy Laboratory;
  \url{https://nwtc.nrel.gov/system/files/TurbSim.pdf}:   2012.

\bibitem{Dunne2016}
Dunne F, Aho J, Pao LY. {Analysis of gain-scheduling implementation for the
  NREL 5-MW turbine blade pitch controller}. In:  {\it Proc. American Control Conference}; 2016\string: 3188--3193

\bibitem{Zalkind2020a}
Zalkind DS, Dall'Anese E, Pao LY. Automatic controller tuning using a
  zeroth-order optimization algorithm. {\it Wind Energy Science} 2020\string;
  5(4)\string: 1579--1600.
\newblock \href {\doibase 10.5194/wes-5-1579-2020} {doi:
  10.5194/wes-5-1579-2020}

\bibitem{Bossanyi2003}
Bossanyi EA. Wind turbine control for load reduction. {\it Wind Energy}
  2003\string; 6(3)\string: 229-244.
\newblock \href {\doibase 10.1002/we.95} {doi: 10.1002/we.95}

\bibitem{Sowento}
Schlipf D. {Controller design and implementation}. ; .
\newblock TTI GmbH - Sowento TGU, \url{https://www.sowento.com/services/}.

\bibitem{Simulink}
{Simulink} . {\it Simulation and Model-Based Design}.
\newblock MathWorks .
\newblock 2020.

\bibitem{Knudsen2013}
Knudsen T, Bak T, Soltani M. Prediction models for wind speed at turbine
  locations in a wind farm. {\it Wind Energy} 2011\string; 14(7)\string:
  877-894.
\newblock \href {\doibase 10.1002/we.491} {doi: 10.1002/we.491}

\bibitem{Simley2016}
Simley E, Pao LY. Evaluation of a wind speed estimator for effective hub-height
  and shear components. {\it Wind Energy} 2016\string; 19(1)\string: 167-184.
\newblock \href {\doibase 10.1002/we.1817} {doi: 10.1002/we.1817}

\bibitem{BarShalom2003}
Bar-Shalom Y, Kirubarajan T, Li XR. {\it Estimation with Applications to Tracking and Navigation}.
\newblock USA: John Wiley \& Sons, Inc. .
\newblock 2002.

\bibitem{Soltani2013}
Soltani MN, Knudsen T, Svenstrup M, Wisniewski R, Brath R, Ortega R, Johnson K.
  {Estimation of rotor effective wind speed: A comparison}. {\it IEEE Transactions on Control Systems Technology} 2013\string; 21(4)\string:
  1155--1167.
\newblock \href {\doibase 10.1109/TCST.2013.2260751} {doi:
  10.1109/TCST.2013.2260751}

\end{thebibliography}




\end{document}